\documentclass[10pt]{article}

\usepackage{graphicx}

\usepackage[all,cmtip]{xy}

\usepackage{setspace}
\usepackage{bbm,youngtab}
\usepackage{amscd,mathrsfs}
\usepackage{amssymb,amsmath,dsfont,amsfonts,amsthm}
\usepackage{euscript,enumerate}
\usepackage{stmaryrd}%for double brackets
\usepackage[cbgreek]{textgreek} %For non-italic greek letters
\usepackage{booktabs}
\usepackage{multirow}
\usepackage{hhline} %For double horizontal lines in tables
\usepackage{color}

\DeclareMathAlphabet{\mathpzc}{OT1}{pzc}{m}{it}

\allowdisplaybreaks %For breaking long equations and putting them on different pages

\usepackage{authblk}
\usepackage{natbib}
\usepackage{color, soul}
\usepackage{enumerate}
\usepackage{empheq}
\usepackage{rotating}
\usepackage{ftnxtra}
\usepackage{fnpos}
\usepackage[titletoc,toc,title]{appendix}
\usepackage{euscript}
\usepackage{graphicx}
\usepackage{epsfig}
\usepackage{epstopdf}
\DeclareGraphicsExtensions{.pdf,.png,.jpg,.eps}
\usepackage{pstool}
\usepackage{upgreek}
\usepackage{mathrsfs}
\usepackage{tikz-cd}

\usepackage{psfrag}

\numberwithin{equation}{section}

\theoremstyle{plain}	
\newtheorem{thm}{Theorem}[section]

\newtheorem*{prop*}{Proposition}
\theoremstyle{definition}	

\newtheorem{remark}[thm]{Remark}

\usepackage{caption}
\usepackage{subcaption}

\usepackage{hyperref}
\hypersetup{colorlinks=true, linkcolor=blue}
\hypersetup{colorlinks=true,citecolor=blue}

\DeclareMathAlphabet{\mathpzc}{OT1}{pzc}{m}{it}

\usepackage{amsmath, amsthm, amssymb}

\usepackage{cleveref}

\DeclarePairedDelimiter\abs{\lvert}{\rvert}
\makeatletter
\newsavebox{\@brx}
\newcommand{\llangle}[1][]{\savebox{\@brx}{\(\m@th{#1\langle}\)}%
  \mathopen{\copy\@brx\mkern2mu\kern-0.9\wd\@brx\usebox{\@brx}}}
\newcommand{\rrangle}[1][]{\savebox{\@brx}{\(\m@th{#1\rangle}\)}%
  \mathclose{\copy\@brx\mkern2mu\kern-0.9\wd\@brx\usebox{\@brx}}}%
\let\oldabs\abs
\def\abs{\@ifstar{\oldabs}{\oldabs*}}
\makeatother

\usepackage{accents}

\usepackage{mathtools}

    {\end{bmatrix}}%

\usepackage[all,cmtip]{xy}

\usepackage{enumitem}

\usepackage{bbold}

\usepackage{yhmath}

\usepackage{amsmath}
\usepackage{arydshln}

\begin{document}
\bibliographystyle{abbrvnat}

\title{\Large{\textbf{\vspace{-.75in}\\
On the Effective Mass of Mechanical Lattices with Microstructure
}}}

\author[1]{Francesco Fedele}
\author[1]{Phanish Suryanarayana}
\author[1,2]{Arash Yavari\thanks{Corresponding author, e-mail: arash.yavari@ce.gatech.edu}}
\affil[1]{\small \textit{School of Civil and Environmental Engineering, Georgia Institute of Technology, Atlanta, GA 30332, USA}}
\affil[2]{\small \textit{The George W. Woodruff School of Mechanical Engineering, Georgia Institute of Technology, Atlanta, GA 30332, USA}}

\maketitle
\thispagestyle{empty}

%-----------------------------
%-----------------------------
\begin{abstract}
\noindent
We present a general formalism for the analysis of mechanical lattices with microstructure using the concept of effective mass. 
We first revisit a classical case of microstructure being modeled by a  spring-interconnected mass-in-mass cell. The frequency-dependent effective mass of the cell is the sum of a static mass and of an added mass, in analogy to that of a swimmer in a fluid. The effective mass is derived using three different methods: momentum equivalence, action equivalence, and dynamic condensation. These methods are generalized to mechanical systems with arbitrary microstructure. As an application, we calculate the effective mass of a $1$D composite lattice with microstructure modeled by a chiral spring-interconnected mass-in-mass cell.  A reduced (condensed) model of the full lattice is then obtained by lumping the microstructure into a single effective mass. A dynamic Bloch analysis is then performed using both the full and reduced lattice models, which give the same spectral results. In particular, the frequency bands follow from the full lattice model by solving a linear eigenvalue problem, or from the reduced lattice model by solving a smaller nonlinear eigenvalue problem. The range of frequencies of negative effective mass falls within the bandgaps of the lattice. Localized modes due to defects in the microstructure have frequencies within the bandgaps, inside the negative-mass range. Defects of the outer, or macro stiffness yield localized modes within each bandgap, but outside the negative-mass range. The proposed formalism can be applied to study the odd properties of coupled micro-macro systems, e.g., active matter.
\end{abstract}

\begin{description}
\item[Keywords:] Effective Mass, Added Mass, Dynamic Mass, Mechanical Lattice, Microstructure, Chiral Solids.
\end{description}

\tableofcontents

%-----------------------------
%-----------------------------
\section{Introduction} \label{Intro}
Solids with microstructure can have dynamic (effective) masses that are significantly different from their static masses \citep{Banerjee2011}. The idea of dynamic  mass has its origins in the works of \citet{Berryman1980} and \citet{Willis1985}, who found that  the effective mass density for composites is not just the average mass density, but is rather dependent on the frequency of excitation.  In these works, the unit cell of the composite is assumed to consist of a heavy mass that is embedded in a very soft matrix. This can be represented by the  classical microstructure model of a hollow rectangular box with mass $M_0$, with another mass $m$ inside it that is connected to the two opposing walls with linear springs of stiffness $k$, also referred to as a mass-in-mass lattice \citep{Huang2009}. Under harmonic excitations with frequency $\omega$, the effective mass has the following expression \citep{Milton2007}\footnote{\citet{Milton2007} considered a unit cell with $n$ cavities. There is a micro mass in each cavity and is connected to the cavity wall by two linear springs. However, the micro masses do not interact with each other. In the case of $n$ cavities their effective mass is
%---------------------
\begin{equation} 
	M_{\rm eff}(\omega)=M_0+\frac{2k}{2k-m\omega^2}\,nm\,.
\end{equation}
%---------------------
} 
%---------------------
\begin{equation}\label{effecmass} 
	M_{\rm eff}(\omega)=M_0+\frac{2k}{2k-m\omega^2}\,m\,,
\end{equation}
%---------------------
which is different from the static mass $M_{\rm eff}(\omega=0)=M_0+m$.
The effective mass is negative when the excitation frequency $\omega$ approaches the natural frequency $\omega_i=2 k/m$ of the system from above, i.e., in the frequency range 
\[
\omega_i<\omega<\omega_i\sqrt{1+\frac{m}{M_0}}\,,
\]
and $M_{\rm eff}(\omega_i\pm\epsilon)=\mp\infty$ and $M_{\rm eff}(\omega_i\sqrt{1+m/M_0})=0$.

The above $1$D problem has been generalized to both $2$D and $3$D by \citet{Milton2007}, in which case the effective mass density becomes matrix-valued. This concept is consistent with the balance of linear momentum---$\operatorname{div}\boldsymbol{\sigma}+\rho\mathbf{b}=\rho\ddot{\mathbf{u}}$, where $\boldsymbol{\sigma}$ is the Cauchy stress, $\mathbf{b}$ is the body force, and $\ddot{\mathbf{u}}$ is the acceleration---which only requires that the inertial force $\rho\ddot{\mathbf{u}}$ must be a vector. Therefore, the mass density can be either a scalar $\rho$ or a second-order tensor $\boldsymbol{\rho}_{\text{eff}}$. 
There  have been numerous  efforts over the past two decades  directed towards the design and analysis of composites with microstructure that display matrix-valued effective mass density \citep{Mei2007,Avila2008,Huang2009,Lai2011}.  The effective mass matrix can have negative eigenvalues, which reduces to the effective mass becoming negative in the case of a $1$D model, as discussed above. In particular, the additional impulse, or momentum generated by the solid shape changes can make the solid move towards the source of force instead of away from it, as if it would have a negative mass.  In such cases, the solid moves and deforms in a non-intuitive way when a force acts on it. If a positive mass is pushed, it accelerates away from the source of the force. A negative mass instead would accelerate towards it. It is worth noting that such negative effective masses are not merely a theoretical construct, but have found experimental realizations \citep{yao2008experimental, yang2013coupled, muhlestein2017experimental}. 

The concept of effective mass is also fundamental in revealing exotic properties of coupled inner-outer (micro-macro) systems, such as active matter or metamaterials~\citep{Vitelli_topologicalmatter}, or mechanical lattices with microstructure. For example, the associated effective lattice, which accounts for the effects of the microstructure on the macrostructure, is characterized by its effective mass. The so-called odd properties such as negative mass or stiffness, no reciprocity of mutual forces, odd viscosity or elasticity~\citep{Vitelli_odd_elasticity} are clearly defined for the effective system, but hidden in the original coupled system. Wave propagation in metamaterials with microstructure is affected in the excitation frequency range of negative mass leading to thickening of the bandgaps, where waves are damped and transmission is prevented \citep{huang2009wave}. It should be mentioned that the discrepancy between the static and dynamic variants of the same physical property is not restricted to mass; for example, it is known that for gradient solids the static and dynamic elastic constants are different \citep{DiVincenzo1986}.\footnote{For elastic transformation cloaking applications the dynamic elastic constants are relevant as was discussed in detail in \citep{Yavari2019, Sozio2021}.}

In this work, using the concept of effective mass, we present a general formalism for the analysis of mechanical lattices with microstructure. 
Specifically, we first consider a $1$D lattice with microstructure. In order to have more than one degree of freedom for each micro or macro element, we assume chiral linear springs that couple longitudinal and torsional degrees of freedom for each element. 
A linear chiral spring is a $1$D  noncentrosymmetric linear elastic solid.
Noncentrosymmetric solids can be modeled in the setting of generalized continuum mechanics and have been studied by many researchers:  \citep{Cheverton1981,Lakes1982,Lakes2001,Sharma2004,liu2012chiral,iecsan2016chiral,Bohmer2020}.
\citet{Papanicolopulos2011} studied chirality in 3D isotropic gradient elasticity under the assumption of small strains. Chirality is controlled by a single material parameter in the fifth-order coupling elasticity tensor.
\citet{auffray2015complete,auffray2017handbook} studied the material symmetries in 2D linear gradient elasticity.
In dimension two, chirality is due to the lack of mirror symmetry, and it affects both the coupling and the second-order elasticity tensors. They showed that there are fourteen symmetry classes, eight of which have isotropic first-order elasticity tensors. 

The remainder of this paper is structured as follows. In \S\ref{Sec:EffectiveMass}, we provide an analogy to the effective mass of a swimmer in a fluid and then revisit the classical spring-interconnected mass-in-mass system. We show that its frequency-dependent effective mass can be derived using three different methods: momentum equivalence, action equivalence, and dynamic condensation of the momentum balance equations. We then generalize the latter methods to mechanical systems with arbitrary microstructure and derive the associated effective mass matrix. In~\S\ref{Sec:1Dcompositelattice}, the proposed formalism is applied to study the reduced~(condensed) model of a $1$D composite lattice with microstructure modeled by a chiral spring-interconnected mass-in-mass cell. The reduced lattice model is obtained by lumping the microstructures into single effective masses and the frequency range of negative mass is determined. A dynamic Bloch analysis is presented in \S\ref{Sec:Blochanalysis} using both the full and the reduced lattice models. In particular, the frequency bands of the lattice are computed. In \S\ref{Sec:defects}, we study the effect of defects on the lattice and the associated localized modes.
Conclusions are given in \S\ref{Sec:Conclusions}.

%-------------------------------------------------------
%-------------------------------------------------------
\section{Effective mass formalism} \label{Sec:EffectiveMass}

In this section, we describe the effective mass formalism for mechanical systems with microstructure. Specifically, in \S\ref{Subsec:Swimmer}, we first motivate the effective mass concept through an analogy with a swimmer in a fluid. Next, in \S\ref{Subsec:1Dprototype}, using three different methods to derive the effective mass, we revisit a system in which the microstructure is modeled as a $1$D spring-interconnected mass-in-mass cell, a case  that has been extensively studied in the literature \citep{Milton2007,Lai2011,Manimala2014,Cveticanin2017,Cveticanin2018}. Finally, in \S\ref{Sec:General}, we extend this framework to generalized mechanical systems with arbitrary microstructure.

%-------------------------------------------------------
%-------------------------------------------------------
\subsection{Swimmer in a fluid analogy} \label{Subsec:Swimmer}

In fluid mechanics, the motion of a swimmer at low Reynolds numbers can be explained in terms of geometric phases~\citep{Saffman1967,Shapere1987}. Swimmers in an ambient fluid can cyclically change their shape to move forward. The coupled swimmer-ambient fluid system conserves the total linear momentum, and since the inertia of the swimmer can be considered to be  negligible, the swimmer velocity is uniquely determined by the geometry of the sequence of its body shapes, which leads to a net translation. Note that only a layer of fluid surrounding the swimmer is altered by its motion and shape deformation. So, in this sense, one can consider the fluid-swimmer interaction as that of a coupled inner-outer (micro-macro) system where the outer component is the swimmer and the  inner component is the portion of the fluid disturbed by the swimmer motion deformation. 

Consider a swimmer of mass $M$ moving in a surrounding ambient inviscid and irrotational fluid of infinite extent. The instantaneous position vector of the swimmer is $\mathbf{X}(t)=\sum_{j=1}^3 X^j\mathbf{e}_j$, where $\mathbf{e}_j$ are the unit vectors of the ambient space $\mathbb{R}^3$. The velocity of the swimmer's center of mass is $\mathbf{U}=\dot{\mathbf{X}}=\sum_{j=1}^3 U^j\mathbf{e}_j$, where the velocity components are defined as $U^j=\dot{X}^j$. Newton's law of motion gives us $\frac{d}{dt}\left(M\mathbf{U}+\mathbf{I}_B\right)=\mathbf{F}_B$, where $\mathbf{F}_B$ is an external force acting on the swimmer and $\mathbf{I}_B$ is the impulse exerted by the surrounding fluid to put the swimmer in motion, or the linear momentum of the fluid: $\mathbf{I}_B=-\rho\int_{\mathcal{S}_B} \Phi\,\mathbf{n}\,\mathrm{d}S$.
Here, $\rho$ is the fluid mass density, $\Phi(\mathbf{x},t)$ is the velocity potential of the fluid flow with velocity field $\mathbf{u}=\nabla \Phi$, $\mathcal{S}_B$ is the boundary surface of the swimmer, and $\mathbf{n}$ is the unit outer normal to the boundary from the body into the fluid. The fluid domain is $\Omega$, where $\mathcal{S}_B$ is its boundary~$\partial \Omega$. The velocity potential $\Phi$ of the fluid flow  is a harmonic function, i.e., $\nabla^2 \Phi=0$ in $\Omega$, with the Neumann boundary condition $\mathbf{u}\cdot \mathbf{e}=\nabla{\Phi}\cdot \mathbf{e}=\mathbf{U}$. Thus, the fluid speed matches with that of the the moving swimmer body, and $\mathbf{e}=\mathbf{U}/\big|\mathbf{U}\big|$  is the unit vector along the direction of motion. 

Since no external forces act on the swimmer, i.e., $\mathbf{F}_B=\mathbf{0}$, the total linear momentum $\mathbf{L}=M\mathbf{U}+\mathbf{I}_B$,
is conserved. 
This implies that 
%-------------------------------------------------------
\begin{equation}\label{totalmomentum}
    M \mathbf{U}+\mathbf{I}_B=\mathbf{0}\,,
\end{equation}
%-------------------------------------------------------
where a zero initial total momentum is assumed, i.e., both the ambient fluid and the swimmer are initially at rest. The impulsive force $\mathbf{I}_B$ is the response of the fluid surrounding the swimmer. The pressure exerted by the thin layer of the surrounding fluid disturbed by the swimmer shape changes varies in such a way that the net fluid pressure force speeds up or slows down the swimmer.
The velocity potential can be decomposed as~\citep{Saffman1967,Shapere1987}
%-------------------------------------------------------
\begin{equation}
    \Phi=\sum_{j=1}^3\Phi_j \dot{X}^j +\sum_{\alpha}\widetilde{\Phi}_{\alpha}\dot{S}^{\alpha}\,,
\end{equation}
%---------
where $\Phi_j$ is the \emph{translation potential} due to the motion of an instantaneously identical rigid body moving at the unit speed along the direction $X^j$. It is a harmonic function satisfying $\nabla^2 \Phi_j=0$ in  $\Omega$ and the Neumann boundary conditions read $\mathbf{u}\cdot \mathbf{e}_j=
\nabla\Phi_j \cdot \mathbf{e}_j=1$.  The \emph{deformation potential} $\widetilde{\Phi}_{\alpha}$ measures the changes of the fluid flow due to a change in shape defined by the deformation displacements $S^{\alpha}$  relative to the rigid body, and $\dot{S}^{\alpha}$
is the speed of deformation,
with $\alpha$ being the index of shape modes.
The potential $\widetilde{\Phi}_{\alpha}$ is a harmonic function satisfying $\nabla^2 \widetilde{\Phi}_{\alpha}=0$ in  $\Omega$ with the Neumann boundary conditions $\mathbf{u}\cdot \mathbf{e}_{\alpha}=\nabla\widetilde{\Phi}_{\alpha}\cdot \mathbf{e}_{\alpha}=1$, where  $\mathbf{e}_{\alpha}$ is the unit vector defining the deformation displacements of the boundary~$\mathcal{S}_B$~\citep{Saffman1967,Shapere1987}.
Then, the impulse components can be written as
%-------------------------------------------------------
\begin{equation}
    (\mathbf{I}_B)_j=M^{(a)}_j \dot{X}^j +\sum_{\alpha}F^{(s)}_{\alpha,j}\,\dot{S}^{\alpha}\,,\quad j=1,2,3\,,
\end{equation}
%-------------------------------------------------------
where
%-------------------------------------------------------
\begin{equation}
   M^{(a)}_j=-\rho\int_{\mathcal{S}_B} \Phi_j \mathbf{n}\cdot \mathbf{e}_j \,dS\,,\qquad  F^{(s)}_{\alpha,j}=-\rho\int_{\mathcal{S}_B} \widetilde{\Phi}_{\alpha} \mathbf{n}\cdot \mathbf{e}_j \,dS\,,\quad j=1,2,3\,. 
\end{equation}
%-------------------------------------------------------
Here, $M^{(a)}_j$ is the added mass and $M^{(a)}_j \dot{X}^j$ is the linear momentum, or impulse generated by the fluid altered by the swimmer moving at the speed $\dot{X}^j$ in the $X^j$ direction. Similarly, $F^{(s)}_{\alpha,j}$ is the linear momentum, or impulse generated by the fluid in the direction $X^j$ by a unit deformation speed of the shape mode $S^{\alpha}$. 
The conservation of total linear momentum in \eqref{totalmomentum} can be written as 
%-------------------------------------------------------
\begin{equation}\label{eq1}
    \left(M+M^{(a)}_j\right) \dot{X}^j +F^{(s)}_{\alpha,j}\,\dot{S}^{\alpha}=0\,,\quad j=1,2,3\,.
\end{equation}
%-------------------------------------------------------
Thus, the swimmer carries with it an added mass $M^{(a)}_j$ of the surrounding fluid. Moreover, as  the swimmer (macro/outer system) changes shape, it alters a layer of the surrounding fluid (micro/inner system). The fluid generates the impulse $F^{(s)}_{\alpha,j}\,\dot{S}_{\alpha}$ in response to the altered pressure distribution around the deforming swimmer. 

Let us now assume that the swimmer can move along the direction $X^1$ only with speed $\dot{X}^1$. We can define an effective mass of the swimmer by equating the total linear momentum of the swimmer-fluid system in \eqref{eq1} to that of an equivalent body of effective mass $M_{\text{eff}}$ moving at the same speed $\dot{X}^1$ as
%-------------------------------------------------------
\begin{equation}\label{effmomentum}
    \left(M+M^{(a)}_1\right) \dot{X}^1 +F^{(s)}_{\alpha}\,\dot{S}_{\alpha}=M_{\text{eff}} \,\dot{X}^1\,,
\end{equation}
%-------------------------------------------------------
from which
%-------------------------------------------------------
\begin{equation}\label{effectivemassfluid}
M_{\text{eff}}=M+M^{(a)}_1 +F^{(s)}_{\alpha,1}\,\frac{\dot{S}^{\alpha}}{\dot{X}^1}\,.
\end{equation}
%-------------------------------------------------------
The effective mass includes the static mass $M$ of the swimmer~(first term), the $\textit{added mass}$~$M^{(a)}_1$~(second term) and an additional $\textit{added mass}$~(third term), which accounts for the effects of the surrounding fluid on the swimmer due to shape changes.
Let us assume a periodic motion, that is $X^1=\hat{X}^1\,\mathrm{e}^{i\omega t}$, and  $S^{\alpha}=\hat{S}^{\alpha}\,\mathrm{e}^{i\omega t}$, where $\hat{X}^1$ and $\hat{S}^{\alpha}$ are Fourier amplitudes. The Fourier transform of the momentum equivalence in \eqref{effmomentum} is $
\left(M+M^{(a)}_1\right) \hat{X}^1 +F^{(s)}_{\alpha}\,\hat{S}_{\alpha}=M_{\text{eff}} \,\hat{X}^1$, and solving for the effective mass yields
%-------------------------------------------------------
\begin{equation}\label{effectivemassfluidFourier}
M_{\text{eff}}=M+M^{(a)}_1 +F^{(s)}_{\alpha,1}\,\frac{\hat{S}^{\alpha}}{\hat{X}^{1}}\,.
\end{equation}
This effective mass follows from an equivalence between the true complex system and an effective system, with the added momentum due to the action of the inner system on the outer system. In particular, the momentum equivalence in \eqref{effmomentum} is the key concept in defining an effective mass, which includes an added mass and the effects of the micro/inner motion (the portion of fluid disturbed by the swimmer) on the macro/outer system (swimmer).
%-----------------------------
%-----------------------------
\begin{figure}[t!]
\centering
\includegraphics[width=0.45\textwidth]{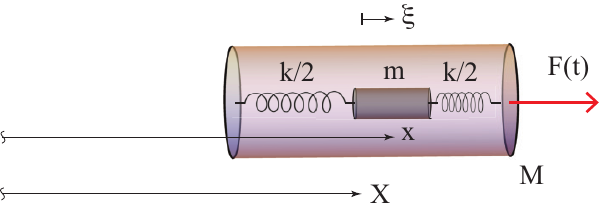}
\vspace*{0.10in}
\caption{A $1$D spring-interconnected mass-in-mass system. The macro element has mass $M$. There is a single micro element inside with mass $m$ that is connected to the macro element by two identical linear springs.} 
\label{Fig:MassInMass}
\end{figure}
%-----------------------------
%-----------------------------

%-------------------------------------------------------
%-------------------------------------------------------
\subsection{$1$D spring-interconnected mass-in-mass microstructure} \label{Subsec:1Dprototype}

We now show that 
mechanical systems with microstructure are the analogue of the fluid-swimmer system. Consider two masses interconnected by two linear springs of stiffness $k/2$, as shown in Fig.~\ref{Fig:MassInMass}. The two springs are in series and their stiffness is equivalent to that of a single spring of stiffness $k$. The dynamical equilibrium equations are  
%-------------------------------------------------------
\begin{equation}
    M \ddot{X} - k (x - X) =F_X(t)\,,
    \quad
     m \ddot{x} + k (x - X) =0\,.
\end{equation}
%-------------------------------------------------------
We distinguish between the outer system of mass $M$ with coordinate $X$ and the inner system of mass $m$ described by $\xi=x-X$. The applied force $F_X(t)$ acts on the outer system. We want to understand how the two systems are dynamically coupled. The outer system is the analogue of a swimmer that advects downstream described by the outer variable $X$. The swimmer changes shape displaying the surrounding fluid, whose motion is  described by the inner variable $\xi$. 
Then, the equations above can be rewritten as
%-------------------------------------------------------
\begin{equation}\label{dyneq}
    M \ddot{X}-k \,\xi =F_X(t)\,,
    \quad
     m \ddot{\xi}+m\ddot{X}+k\, \xi =0\,.
\end{equation}
%------------------------------------------------------
Adding the two equations one obtains 
%-------------------------------------------------------
\begin{equation} \label{applied-force}
    (M +m)\ddot{X} + m\,\ddot{\xi}= F_X(t)\,,
\end{equation}
%------------------------------------------------------
which can be written as $\frac{d\mathsf{A}}{dt}= F_X(t)$,
where we have defined the total linear momentum as 
%-------------------------------------------------------
\begin{equation} \label{lin-momentum}
   \mathsf{A}=(M +m)\dot{X} +m \dot{\xi}\,.
\end{equation}
%------------------------------------------------------
Thus,
%-------------------------------------------------------
\begin{equation}
    \mathsf{A}=\mathsf{A}_0 +\int_0^{t} F_X(\tau)\,d\tau\,,
\end{equation}
%------------------------------------------------------
where $\mathsf{A}_0$ is the initial linear momentum. 
%-------------------------------------------------------
Therefore, the motion of the outer system is given by
%-------------------------------------------------------
\begin{equation}
   \dot{X} =\frac{\mathsf{A}_0}{M +m} -\frac{m}{M+m}\,\dot{\xi}+ \int_0^{t} \frac{ F_X(\tau) }{M+m}\,d\tau\,.
\end{equation}
%------------------------------------------------------
Thus, the motion of the outer system depends on the linear momentum of the entire system in the absence of external forces ($F_X=0$) and internal motion, i.e., $\xi=0$. 

%------------------------------------------------------
%------------------------------------------------------
\paragraph{Momentum equivalence.}
The inner motion,  or shape deformation,~$\dot{\xi}$ contributes to an added momentum, which can slow down or speed up the entire system. The momentum depends also on the total impulse generated by the external forces. 
Thus, one can define an effective mass of an equivalent mass-spring system as $M_{\text{eff}} \ddot{X}=F_{\text{eff}}(t)$, which describes the outer system and accounts for the momentum added by the coupling with the inner system (dynamic condensation). Its momentum is 
%-------------------------------------------------------
\begin{equation} \label{eff-lin-momentum}   \mathsf{A}_{\text{eff}}=M_{\text{eff}}\dot{X}\,,
\end{equation}
%------------------------------------------------------
and
%-------------------------------------------------------
\begin{equation}
    \dot{X} = \frac{\mathsf{A}_{\text{eff}}(t=0)}{M_{\text{eff}}} + \int_0^{t} \frac{ F_{\text{eff}}(\tau)}{M_{\text{eff}}}\,d\tau\,.
\end{equation}
%------------------------------------------------------
Equating the momenta of the two systems, i.e., $\mathsf{A}_{\text{eff}}=\mathsf{A}$, from \eqref{lin-momentum} and \eqref{eff-lin-momentum} we have
%-------------------------------------------------------
\begin{equation}\label{momentum}
    M_{\text{eff}} \dot{X}=(M+m)\dot{X}+m\,\dot{\xi}\,,
\end{equation}
%------------------------------------------------------
and the effective mass is given as
%-------------------------------------------------------
\begin{equation}
    M_{\text{eff}}=(M+m) + m \,\frac{\dot{\xi}}{\dot{X}}\,,
\end{equation}
%------------------------------------------------------
which is similar to the effective mass of a swimmer in \eqref{effectivemassfluid}. Indeed, the first two terms together is the total mass of the system as a rigid body (static mass). This includes the $\emph{added mass}$ $m$ of the inner system~(fluid), which is dragged by the outer system (swimmer). The third term $m\,\dot{\xi}/\dot{X}$ is an additional $\emph{added mass}$ due to the internal deformations of the inner system in analogy with the impulse induced by the swimmer shape change in \eqref{effectivemassfluid}. 
Comparing \eqref{applied-force} and the time derivative of \eqref{momentum}, one concludes that the equivalent force is $F_{\text{eff}}(t)=F_X(t)$.

Let us assume that the coupled system is subject to periodic forcing. 
Then the Fourier modes of displacements and forces are $X=\hat{X}\,\mathrm{e}^{\omega t}$, $\xi=\hat{\xi}\,\mathrm{e}^{\omega t}$, and $F_X=\hat{F}_X\,\mathrm{e}^{\omega t}$, where $(\hat{X},\hat{\xi},\hat{F})$ are complex amplitudes. Then, from \eqref{momentum} equating the momenta yields $M_{\text{eff}}\,i\omega \hat{X}=(M+m)\,i\omega \hat{X} + m\,i\omega \hat{\xi}$, from which
%-------------------------------------------------------
\begin{equation}\label{effectivemass}
    M_{\text{eff}}=(m+M) + m\,\frac{\hat{\xi}}{\hat{X}}\,.
\end{equation}
%------------------------------------------------------
From \eqref{dyneq}, the dynamical equation of the inner motion transforms in Fourier space to $-m\,\omega^2\hat{\xi}  -m\omega^2\hat{X}+ k\, \hat{\xi}=0$,
and hence, 
%-------------------------------------------------------
\begin{equation}\label{dyneq3}
    \hat{\xi}= \frac{m \omega^2}{k -m \omega^2}\,\hat{X}\,.
\end{equation}
%------------------------------------------------------
The effective mass in \eqref{effectivemass} is now simplified to read
%-------------------------------------------------------
\begin{equation}\label{effectivemass2}
    M_{\text{eff}}(\omega)=(M+m) + m\,\frac{m \omega^2}{k -m \omega^2}=M + m\,\frac{k}{k-m \omega^2}\,,
\end{equation}
%------------------------------------------------------
and the static mass is $M_{\text{eff}}(\omega=0)=M+m$.  Note that the effective mass is negative when the
excitation frequency $\omega$ approaches the natural frequency $\omega_i = k/m$ of the inner system from above in the frequency range
$\omega_i <\omega < \omega_i\sqrt{1+m/M}$. Here, $M_{\text{eff}}(\omega_i\pm\epsilon)=\mp\infty$, and $M_{\text{eff}}(\omega_i\sqrt{1+m/M})=0$. 

\remark{An alternate strategy to the above is matching the inertial forces $\dot{\mathsf{A}}_e=\dot{\mathsf{A}}$, arriving at an alternative form for the effective mass:
%-------------------------------------------------------
\begin{equation}
    M_{\text{eff}}=(M+m)+ m\, \frac{\ddot{\xi}}{\ddot{X}}\,,
\end{equation}
%------------------------------------------------------
where the second term $m \ddot{\xi}/\ddot{X}$ is an additional $\textit{added mass}$ due to the internal inertia of the system. For a periodic motion, it will lead to the same effective mass given in \eqref{effectivemass2}. }

\paragraph{Action equivalence.} The Lagrangian of the two-mass-spring system considered above is written as $\mathsf{L}=\mathsf{K}-\mathsf{P}-\mathsf{W}$, where the kinetic energy $\mathsf{K}$, the potential energy $\mathsf{P}$, and the work $\mathsf{W}$ done by the external forces are given by
%------------------------------------------------------
\begin{equation}\label{KPW}
 \mathsf{K}=\frac{1}{2} M \dot{X}^2 +\frac{1}{2} m \dot{x}^2\,,\qquad \mathsf{P}=\frac{1}{2} k \xi ^2\,,\qquad \mathsf{W}=F_X(t) X(t)\,. 
\end{equation}
%------------------------------------------------------
The Lagrangian density of the equivalent system with effective mass $M_{\text{eff}}$ and subject to the effective force $F_{\text{eff}}$ is
%------------------------------------------------------
\begin{equation}\label{effLagrang2}
 	\mathsf{L}_{\text{eff}}=\frac{1}{2} M_{\text{eff}} \dot{X}^2 - F_{\text{eff}}(t) X(t)\,,
\end{equation}
%-------------------------------------------------
where the equivalence is meant in the sense that the two Lagrangians are the same on average, that is  $\overline{\mathsf{L}}=\overline{\mathsf{L}}_{\text{eff}}$, where $\overline{f}=\lim_{T\rightarrow\infty}\frac{1}{T}\int_0^T f(t)\mathrm{d}t$ is the time average of $f$.
Consider a periodic force with frequency $\omega$ given by 
%-------------------------------------------------
\begin{equation}
	F_{X}(t)=|\hat{F}_{X}|\cos(\omega t + \phi_{X})
	=\frac{1}{2}\hat{F}_{X}\mathrm{e}^{i\omega t} + \text{c.c.}\,,
\end{equation}
%-------------------------------------------------
where the complex amplitude is defined as $\hat{F}_{X}=|\hat{F}_{X}|\mathrm{e}^{i \phi_{X}}$, and \emph{c.c.} denotes complex conjugate. The average external work in \eqref{KPW}$_3$ is given by
%-------------------------------------------------
\begin{equation}\label{Work2}
    \overline{\mathsf{W}}=\frac{1}{4}\hat{F}_{X}^{\dagger} \hat{X}+\text{c.c.}\,,
\end{equation}
%------------------------------------------------------
where $\hat{X}$ is the Fourier amplitude of $X$ and the the operator $\dagger$ is the complex conjugate transpose. The time average of the kinetic energy $\mathsf{K}$ in \eqref{KPW}$_1$ follows as
%------------------------------------------------------
\begin{equation}\label{kineticenergy2}
\begin{aligned}
    \overline{\mathsf{K}} &=\frac{1}{8}\omega^2 M |\hat{X}|^2 +\frac{1}{8}\omega^2 m |\hat{\xi}+\hat{X}|^2+\text{c.c.}=\frac{1}{8}\omega^2 \left(M+m (1+\gamma)^2\right)|\hat{X}|^2+\text{c.c.}\,,
\end{aligned}
\end{equation}
%-------------------------------------------------
where \eqref{dyneq3} was used to solve for the inner displacement $\hat{\xi}=\gamma \hat{X}$, and $\gamma=m \omega^2/(k -m \omega^2)$. Similarly, the time average of the potential energy  $\mathsf{P}$ in \eqref{KPW}$_2$ reads
%-------------------------------------------------
\begin{equation}\label{Potentialeff2}
\begin{aligned}
   \overline{\mathsf{P}} &=\frac{1}{8}k |\hat{\xi}|^2 +\text{c.c.}=\frac{1}{8} k \gamma^2|\hat{X}|^2 + \text{c.c.}\,.
\end{aligned}
\end{equation}
%-------------------------------------------------
The time-average of the Lagrangian $\mathsf{L}$ of the two-mass-spring system simplifies to read
%-------------------------------------------------
\begin{equation}\label{avLagr}
   \overline{\mathsf{L}}=\overline{\mathsf{K}}-\overline{\mathsf{P}}-\overline{\mathsf{W}} =\frac{1}{8}\omega^2 \left(M + m (1+\gamma)^2 - \frac{k}{\omega^2}\gamma^2\right) |\hat{X}|^2 -\frac{1}{4}\hat{F}_{X}^{\dagger} \hat{X} + \text{c.c.}\,,
\end{equation}
%-------------------------------------------------
and that of the effective-mass system in \eqref{effLagrang2} is given by
%-------------------------------------------------
\begin{equation}\label{Lagranlump5}
\overline{\mathsf{L}}_{\text{eff}}=\frac{1}{8} \omega^2 M_{\text{eff}} |\hat{X}|^2-\frac{1}{4}\hat{F}_{\text{eff}}^{\dagger}\hat{X} + \text{c.c.} \,.
\end{equation}
%-------------------------------------------------
Equating the two averaged Lagrangians, i.e., $\overline{\mathsf{L}}=\overline{\mathsf{L}}_{\text{eff}}$, yields $F_{\text{eff}}=F_X$ and the effective mass:
%-------------------------------------------------
\begin{equation} \label{Eq:Meff1DClassAc}
   M_{\text{eff}}=M + m (1+\gamma)^2 - \frac{k}{\omega^2}\gamma^2=M + m\frac{k}{k-m \omega^2}\,. 
\end{equation}

%------------------------------------------------------
\paragraph{Dynamic condensation.}
The effective mass can also be derived by applying the standard approach of matrix condensation. From \eqref{dyneq}, the dynamical equation of the outer motion transforms in the Fourier space to $-M\,\omega^2\hat{X} - k \,\hat{\xi}=\hat{F}_X$, and plugging in the expression of $\hat{\xi}$ of \eqref{dyneq3} one gets
%-------------------------------------------------------
\begin{equation}\label{dyneq8}
     -M\,\omega^2\hat{X} - k \,\frac{m \omega^2}{k -m \omega^2}\hat{X}=\hat{F}_X\,,
\end{equation}
%------------------------------------------------------
which can be written as $-\omega^2\left(M +m\frac{k}{k -m} \right)\hat{X}=\hat{F}_X$.
Thus, we can define the effective mass as
%-------------------------------------------------------
\begin{equation}\label{meff10}
     M_{\text{eff}}=M +m\, \frac{k}{k -m \omega^2}\,. 
\end{equation}
%------------------------------------------------------

\remark{The effective mass obtained using the three different approaches, namely, momentum equivalence \eqref{effectivemass2}, action equivalence \eqref{Eq:Meff1DClassAc}, and dynamic condensation \eqref{meff10}, are identical and coincide with the established results in the literature \citep{Milton2007,Banerjee2011}}

%------------------------------------------------------
%------------------------------------------------------
\subsection{General mechanical systems with arbitrary microstructure} \label{Sec:General}

We now formulate the concept of added mass for a general mechanical system with arbitrary microstructure. 
In so doing, we derive the associated effective mass matrix  and force vector using the three methods described above, namely, momentum equivalence, action equivalence, and dynamic condensation. 

Consider a mechanical system with miscrostructure in which the inner (micro) and outer (macro) systems are coupled according to the following dynamical equations:
%------------------------------------------------------
\begin{equation}\label{IOmatrixsys}
    \begin{bmatrix}
    \mathbf{M}_{\text{OO}} & \mathbf{0}\\
    \mathbf{0} & \mathbf{M}_{\text{II}}
    \end{bmatrix}
    \begin{bmatrix}
    \ddot{\mathbf{X}}_{\text{O}}\\
    \ddot{\mathbf{X}}_{\text{I}}
    \end{bmatrix} 
    +\begin{bmatrix}
    \mathbf{K}_{\text{OO}} & \mathbf{K}_{\text{OI}}\\
    \mathbf{K}_{\text{IO}} & \mathbf{K}_{\text{II}}
    \end{bmatrix}
    \begin{bmatrix}
    \mathbf{X}_{\text{O}}\\
    \mathbf{X}_{\text{I}}
    \end{bmatrix}=
    \begin{bmatrix}
    \mathbf{F}_{\text{O}}(t)\\
    \mathbf{F}_{\text{I}}(t)
    \end{bmatrix}\,.
\end{equation}
%------------------------------------------------------
The outer (macro) system is described by the generalized displacement vector $\mathbf{X}_{\text{O}}\in\mathbb{R}^{N_{\text{O}}}$, and inner (micro) system by the generalized displacement vector $\mathbf{X}_{\text{I}}\in\mathbb{R}^{N_{\text{I}}}$. The respective stiffness matrices are the matrices $\mathbf{K}_{\text{OO}}\in\mathbb{R}^{N_{\text{O}}\times N_{\text{O}}}$, and $\mathbf{K}_{\text{II}}\in\mathbb{R}^{N_{\text{I}}\times N_{\text{I}}}$. The mechanical coupling between the two systems is described by the matrix $\mathbf{K}_{\text{OI}}\in\mathbb{R}^{N_{\text{O}}\times N_{\text{I}}}$, and $\mathbf{K}_{\text{IO}}=\mathbf{K}_{\text{OI}}^\mathsf{T}$, where $\mathsf{T}$ denotes matrix transposition. Both systems are subject to forcing via the force vectors $\mathbf{F}_{\text{O}}\in\mathbb{R}^{N_{\text{O}}}$, and $\mathbf{F}_{\text{I}}\in\mathbb{R}^{N_{\text{I}}}$. 

The dynamical equations in \eqref{IOmatrixsys} can be expanded as two coupled equations for the inner and outer systems as 
%------------------------------------------------------
\begin{equation}\label{maeq}
\begin{aligned}
    \mathbf{M}_{\text{OO}}\ddot{\mathbf{X}}_{\text{O}} + \mathbf{K}_{\text{OO}}\mathbf{X}_{\text{O}}  + \mathbf{K}_{\text{OI}}\mathbf{X}_{\text{I}} &=\mathbf{F}_{\text{O}}(t)\,, \\
    \mathbf{M}_{\text{II}}\ddot{\mathbf{X}}_{\text{I}} + \mathbf{K}_{\text{IO}}\mathbf{X}_{\text{O}}  + \mathbf{K}_{\text{II}}\mathbf{X}_{\text{I}} &=\mathbf{F}_{\text{I}}(t)\,.
\end{aligned}
\end{equation}
%------------------------------------------------------
These two equations are the starting point for deriving an effective mass matrix and force vector for the outer system that accounts for the added momentum of the inner system.

%------------------------------------------------------

%-------------------------------------------------
%-------------------------------------------------
\subsubsection{Momentum equivalence}

Adding the two dynamical equations in \eqref{maeq}  for the inner-outer system, and integrating over time gives the conservation of the total momentum\footnote{Depending on the mechanical system this can be linear momentum, angular momentum, or both. In general, by ``momentum" we mean linear and angular momenta.} 
%------------------------------------------------------
\begin{equation}\label{maeq2}
   \boldsymbol{\mathsf{A}}(t)+\int_{t_0}^{t}\left[ (\mathbf{K}_{\text{OO}}+\mathbf{K}_{\text{IO}})\mathbf{X}_{\text{O}}  + (\mathbf{K}_{\text{II}}+\mathbf{K}_{\text{OI}})\mathbf{X}_{\text{I}} \right]\,d\tau -  \int_{t_0}^{t}\left(\mathbf{F}_{\text{O}} + \mathbf{F}_{\text{I}}\right)\,d\tau=\boldsymbol{\mathsf{A}}(t_0)\,,
\end{equation}
%------------------------------------------------------
where the total momentum vector is defined as
\begin{equation}\label{mom4}
   \boldsymbol{\mathsf{A}}(t)=\mathbf{M}_{\text{OO}}\dot{\mathbf{X}}_{\text{O}}+\mathbf{M}_{\text{II}}\dot{\mathbf{X}}_{\text{I}}\,,
\end{equation}
%------------------------------------------------------
and $\dot{\mathbf{X}}=d \mathbf{X}/d t$ denotes the time derivative of $\mathbf{X}$. The time integrals are the impulses of the macro and micro (generalized) forces.\footnote{By~\emph{force} we mean either a force or a moment, and hence, a generalized force.} Let us now define an effective macro system such that
%------------------------------------------------------
\begin{equation}
    \mathbf{M}_{\text{eff}}\, \ddot{\mathbf{X}}_{\text{O}}=\mathbf{F}_{\text{eff}}\,.
\end{equation}
%------------------------------------------------------
The effective momentum vector is defined as $\boldsymbol{\mathsf{A}}_{\text{eff}}=\mathbf{M}_{\text{eff}} \,\dot{\mathbf{X}}_{\text{O}}$ and 
%------------------------------------------------------
\begin{equation}
    \boldsymbol{\mathsf{A}}_{\text{eff}}(t)-\int_{t_0}^{t} \mathbf{F}_{\text{eff}}\,d\tau=\boldsymbol{\mathsf{A}}_{\text{eff}}(t_0)\,,
\end{equation}
%------------------------------------------------------
where the time integral is the impulse  of the effective force vector.
Equating the initial momenta $\boldsymbol{\mathsf{A}}(t_0)=\boldsymbol{\mathsf{A}}_{\text{eff}}(t_0)$ yields 
%------------------------------------------------------
\begin{equation}\label{maeq5}
\begin{aligned}
   \mathbf{M}_{\text{eff}} \dot{\mathbf{X}}_{\text{O}}-\int_{t_0}^{t} \mathbf{F}_{\text{eff}}\,d\tau &=\mathbf{M}_{\text{OO}} \dot{\mathbf{X}}_{\text{O}}+\mathbf{M}_{\text{II}} \dot{\mathbf{X}}_I+\int_{t_0}^{t}\left[ (\mathbf{K}_{\text{OO}}+\mathbf{K}_{\text{IO}})\mathbf{X}_{\text{O}}  + (\mathbf{K}_{\text{II}}+\mathbf{K}_{\text{OI}})\mathbf{X}_{\text{I}} \right]\,d\tau \\
   & \quad -  \int_{t_0}^{t}\left(\mathbf{F}_{\text{O}} + \mathbf{F}_{\text{I}}\right)\,d\tau\,.
\end{aligned}
\end{equation}
%------------------------------------------------------
To formulate an effective mass, we will take the Laplace transform of the above momentum equivalence equation. Consider $t_0=0$ and Laplace transforming \eqref{maeq5} yields
%------------------------------------------------------
\begin{equation}\label{LaplaceT}
\begin{aligned}
   \mathbf{M}_{\text{eff}}(s) \left[s\widetilde{\mathbf{X}}_{\text{O}}(s)- \mathbf{X}_{\text{O}}(0)\right]
   -\frac{\widetilde{\mathbf{F}}_{\text{eff}}(s)}{s} &=
   \mathbf{M}_{\text{OO}} \left[s\widetilde{\mathbf{X}}_{\text{O}}(s)-\mathbf{X}_{\text{O}}(0)\right]
   +\mathbf{M}_{\text{II}} \left[s\widetilde{\mathbf{X}}_{\text{I}}(s)-\mathbf{X}_\text{I}(0)\right]\\
   & \quad +\frac{ (\mathbf{K}_{\text{OO}}+\mathbf{K}_{\text{IO}})\widetilde{\mathbf{X}}_{\text{O}}(s)  + (\mathbf{K}_{\text{II}}+\mathbf{K}_{\text{OI}})\widetilde{\mathbf{X}}_{\text{I}}(s) }{s} \\
   & \quad -\frac{\widetilde{\mathbf{F}}_{\text{O}}(s) + \widetilde{\mathbf{F}}_\text{I}(s)}{s}\,,
   \end{aligned}
\end{equation}
%------------------------------------------------------
where  $\big(\widetilde{\mathbf{X}}_{\text{O}}(s),\widetilde{\mathbf{X}}_{\text{I}}(s)\big)$  are the Laplace transforms of the outer and inner variables $(\mathbf{X}_{\text{O}},\mathbf{X}_{\text{I}})$, which follow from Laplace transforming the dynamical equations in \eqref{maeq}. Assuming the initial conditions $\mathbf{X}_{\text{O}}(0)=\mathbf{X}_{\text{I}}(0)=\mathbf{X}_{\text{eff}}(0)=\mathbf{0}$, we have
%------------------------------------------------------
\begin{equation}\label{LaplaceT3}
\begin{aligned}
   s\mathbf{M}_{\text{eff}} \widetilde{\mathbf{X}}_{\text{O}}(s)-\frac{\widetilde{\mathbf{F}}_{\text{eff}}(s)}{s} &=
   \left(s \mathbf{M}_{\text{OO}}+\frac{\mathbf{K}_{\text{OO}}+\mathbf{K}_{\text{IO}}}{s}\right)\widetilde{\mathbf{X}}_{\text{O}}(s)\\
   & \quad +\left(s\mathbf{M}_{\text{II}} +\frac{\mathbf{K}_{\text{II}}+\mathbf{K}_{\text{OI}}}{s}\right)\widetilde{\mathbf{X}}_{\text{I}}(s)-\frac{\widetilde{\mathbf{F}}_{\text{O}}(s) + \widetilde{\mathbf{F}}_{\text{I}}(s)}{s}\,.
   \end{aligned}
\end{equation}
%------------------------------------------------------
The Laplace transform $\widetilde{\mathbf{X}}_{\text{I}}(s)$ of the inner variable follows from the time-domain equation~\eqref{maeq}$_2$ as
%------------------------------------------------------
\begin{equation}
\mathbf{M}_{\text{II}}\left[s^2\widetilde{\mathbf{X}}_{\text{I}}-s\mathbf{X}_{\text{I}}(0)-\dot{\mathbf{X}}_{\text{I}}(0)\right]+ \mathbf{K}_{\text{IO}}\widetilde{\mathbf{X}}_{\text{O}} +\mathbf{K}_{\text{II}}\widetilde{\mathbf{X}}_{\text{I}}=\widetilde{\mathbf{F}}_{\text{I}}(s)\,.
\end{equation}
%------------------------------------------------------
Assume the initial conditions $\mathbf{X}_{\text{I}}(0)=\dot{\mathbf{X}}_{\text{I}}(0)=\mathbf{0}$. Hence, $(s^2\mathbf{M}_{\text{II}}+\mathbf{K}_{\text{II}})\widetilde{\mathbf{X}}_{\text{I}}=-\mathbf{K}_{\text{IO}}\widetilde{\mathbf{X}}_{\text{O}}+\widetilde{\mathbf{F}}_{\text{I}}$, from which
%------------------------------------------------------
\begin{equation}\label{eqXILaplaceT3}
    \widetilde{\mathbf{X}}_{\text{I}}=(s^2\mathbf{M}_{\text{II}}+\mathbf{K}_{\text{II}})^{-1}\left(\widetilde{\mathbf{F}}_{\text{I}}-\mathbf{K}_{\text{IO}}\widetilde{\mathbf{X}}_{\text{O}}\right)\,.
\end{equation}
%------------------------------------------------------
Then, from \eqref{LaplaceT3} one obtains
%------------------------------------------------------
\begin{equation} \label{LaplaceT6}
\begin{aligned}
   s\mathbf{M}_{\text{eff}}(s)\widetilde{\mathbf{X}}_{\text{O}}(s)-\frac{\widetilde{\mathbf{F}}_{\text{eff}}(s)}{s} &=
   \left(s \mathbf{M}_{\text{OO}}+\frac{\mathbf{K}_{\text{OO}}+\mathbf{K}_{\text{IO}}}{s}\right)\widetilde{\mathbf{X}}_{\text{O}}(s)\\
   & \quad +\frac{(s^2\mathbf{M}_{\text{II}} +\mathbf{K}_{\text{II}})+\mathbf{K}_{\text{OI}}}{s}(s^2\mathbf{M}_{\text{II}}+\mathbf{K}_{\text{II}})^{-1}\left(\widetilde{\mathbf{F}}_{\text{I}}-\mathbf{K}_{\text{IO}}\widetilde{\mathbf{X}}_{\text{O}}\right) -\frac{\widetilde{\mathbf{F}}_{\text{O}}(s) + \widetilde{\mathbf{F}}_{\text{I}}(s)}{s} \\
    &=
   \left(s \mathbf{M}_{\text{OO}}+\frac{\mathbf{K}_{\text{OO}}-\mathbf{K}_{\text{OI}}(s^2\mathbf{M}_{\text{II}}+\mathbf{K}_{\text{II}})^{-1}\mathbf{K}_{\text{IO}}}{s}\right)\widetilde{\mathbf{X}}_{\text{O}}\\
   & \quad -\frac{\widetilde{\mathbf{F}}_O-\mathbf{K}_{\text{OI}}(s^2\mathbf{M}_{\text{II}}+\mathbf{K}_{\text{II}})^{-1}\widetilde{\mathbf{F}}_{\text{I}}}{s}\,.
   \end{aligned}
\end{equation}
%------------------------------------------------------
Therefore, the effective mass matrix is written as
%------------------------------------------------------
\begin{equation}\label{massLaplace}
   \mathbf{M}_{\text{eff}}(s)=
   \mathbf{M}_{\text{OO}}+\frac{\mathbf{K}_{\text{OO}}-\mathbf{K}_{\text{OI}}(s^2\mathbf{M}_{\text{II}}+\mathbf{K}_{\text{II}})^{-1}\mathbf{K}_{\text{IO}}}{s^2}\,.
\end{equation}
%------------------------------------------------------
The effective force vector reads
%------------------------------------------------------
\begin{equation}\label{forcingLaplace}
   \widetilde{\mathbf{F}}_{\text{eff}}(s)=
   \widetilde{\mathbf{F}}_{\text{O}}-\mathbf{K}_{\text{OI}}(s^2\mathbf{M}_{\text{II}}+\mathbf{K}_{\text{II}})^{-1}\widetilde{\mathbf{F}}_{\text{I}}\,.
\end{equation}
%------------------------------------------------------
Then, $\widetilde{\mathbf{X}}_{\text{O}}(s)=\widetilde{\mathbf{H}}_{\text{eff}}(s)\,\widetilde{\mathbf{F}}_{\text{eff}}(s)$, where we have defined the transfer function $\widetilde{\mathbf{H}}_{\text{eff}}(s)=
\mathbf{M}_{\text{eff}}^{-1}(s)$, which is the Laplace transform of the impulsive response of the condensed outer system.
 
The Fourier Transform $\hat{\mathbf{X}}(\omega)$ of $\mathbf{X}(t)$ follows from the Laplace transform $\widetilde{\mathbf{X}}(s)$ by setting $s=i\omega$, i.e., $\hat{\mathbf{X}}(\omega)=\widetilde{\mathbf{X}}(s=i\omega)$. Then, the harmonic expression of the effective mass matrix is written as
%------------------------------------------------------
\begin{equation}\label{massFourier}
   \mathbf{M}_{\text{eff}}(\omega)=\mathbf{M}_{\text{eff}}(s=i\omega)=
   \mathbf{M}_{\text{static}}+\mathbf{M}_{\text{added}}(\omega)\,,
\end{equation}
%-----------------------------------------------------
where 
%------------------------------------------------------
\begin{equation}\label{massFourier}
  \mathbf{M}_{\text{static}}=
   \mathbf{M}_{\text{OO}}\,,\qquad
   \mathbf{M}_{\text{added}}(\omega)=-\frac{1}{\omega^2}\left[\mathbf{K}_{\text{OO}}+\mathbf{K}_{\text{OI}}(\omega^2\mathbf{M}_{\text{II}}-\mathbf{K}_{\text{II}})^{-1}\mathbf{K}_{\text{IO}}\right]\,.
\end{equation}
%-----------------------------------------------------
The effective mass matrix includes the static mass $\mathbf{M}_{\text{static}}$ of the outer system and a frequency-dependent \textit{added mass} $ \mathbf{M}_{\text{added}}(\omega)$ that accounts for the effects of the micro-momentum on the outer system in analogy with the effective mass of a swimmer in a fluid~(see \S\ref{Subsec:Swimmer}). 
The effective force vector is given by
%------------------------------------------------------
\begin{equation}\label{forcingFourier}
   \hat{\mathbf{F}}_{\text{eff}}(\omega)=\widetilde{\mathbf{F}}_{\text{eff}}(s=i\omega)=
   \widetilde{\mathbf{F}}_{\text{O}}+\mathbf{K}_{\text{OI}}(\omega^2\mathbf{M}_{\text{II}}-\mathbf{K}_{\text{II}})^{-1}\widetilde{\mathbf{F}}_{\text{I}}\,.
\end{equation}
%------------------------------------------------------

%------------------------------------------------------
%------------------------------------------------------
\subsubsection{Action equivalence}

The effective mass includes both the effects of the momentum and the elastic forces of the inner system, which are coupled. Let us consider the  Lagrangian of the coupled system 
%------------------------------------------------------
\begin{equation}\label{Lagrang}
    \mathsf{L}=\mathsf{K}- \mathsf{P} -\mathsf{W}\,,
\end{equation}
%-------------------------------------------------
where the kinetic and potential energies are defined as 
%------------------------------------------------------
\begin{equation}\label{kineticenergy}
	\mathsf{K}=\frac{1}{2}\dot{\mathbf{X}}_{\text{O}}^{\mathsf{T}}\mathbf{M}_{\text{OO}}
	\dot{\mathbf{X}}_{\text{O}} +\frac{1}{2} \dot{\mathbf{X}}_{\text{I}}^{\mathsf{T}}\mathbf{M}_{\text{II}}
	\dot{\mathbf{X}}_{\text{I}}\,,\qquad \mathsf{P}=\frac{1}{2}\left(\mathbf{X}_{\text{O}}^{\mathsf{T}}
	\mathbf{K}_{\text{OO}}\mathbf{X}_{\text{O}}+\mathbf{X}_{\text{I}}^{\mathsf{T}}
	\mathbf{K}_{\text{II}}\mathbf{X}_{\text{I}}  
	+ 2\mathbf{X}_{\text{O}}^{\mathsf{T}}\mathbf{K}_{\text{OI}}\mathbf{X}_{\text{I}}\right)\,,
\end{equation}
%-------------------------------------------------
and the work done by the external forces is defined as $\mathsf{W}=\mathbf{F}_{\text{O}}^{\mathsf{T}}(t) \mathbf{X}_{\text{O}}(t) + \mathbf{F}_{\text{I}}^{\mathsf{T}}(t) \mathbf{X}_{\text{I}}(t)$.
Consider the Lagrangian of the equivalent system with the effective mass $\mathbf{M}_{\text{eff}}$ and potential energy $\mathsf{P}_{\text{eff}}$ subject to the effective force $\mathbf{F}_{\text{eff}}$ is given by
%------------------------------------------------------
\begin{equation}\label{effLagrang}
 	\mathsf{L}_{\text{eff}}=\mathsf{K}_{\text{eff}}-\mathsf{P}_{\text{eff}}-\mathsf{W}_{\text{eff}},\qquad 
	\mathsf{K}_{\text{eff}}=\frac{1}{2}\dot{\mathbf{X}}_{\text{O}}^{\mathsf{T}}\mathbf{M}_{\text{eff}}
	\dot{\mathbf{X}}_{\text{O}},\qquad \mathsf{W}_{\text{eff}}
	= \mathbf{F}_{\text{eff}}^{\mathsf{T}}\mathbf{X}_{\text{O}}\,,
\end{equation}
%-------------------------------------------------
where the equivalence is meant in the sense that the two Lagrangians are the same on average, that is  $\overline{\mathsf{L}}=\overline{\mathsf{L}}_{\text{eff}}$, or more explicitly
%------------------------------------------------------
\begin{equation}\label{Lagrangmean}
	\overline{\mathsf{K}}-\overline{\mathsf{P}}-\overline{\mathsf{W}}
	=\overline{\mathsf{K}}_{\text{eff}}-\overline{\mathsf{P}}_{\text{eff}}-\overline{\mathsf{W}}_{\text{eff}}\,,
\end{equation}
%-------------------------------------------------
where $\overline{f}=\lim_{T\rightarrow\infty}\frac{1}{T}\int_0^T f(t)\mathrm{d}t$ is the time average of $f$.
Consider periodic force vectors with frequency $\omega$ given by 
%-------------------------------------------------
\begin{equation}
	\mathbf{F}_{\text{O}}(t)=|\hat{\mathbf{F}}_{\text{O}}|\cos(\omega t + \phi_{\text{O}})
	=\frac{1}{2}\hat{\mathbf{F}}_{\text{O}}\mathrm{e}^{i\omega t} + \text{c.c.},\qquad\mathbf{F}_{\text{I}}(t)
	=|\hat{\mathbf{F}}_{\text{I}}|\cos(\omega t + \phi_{\text{I}})
	=\frac{1}{2}\hat{\mathbf{F}}_{\text{I}}\mathrm{e}^{i\omega t} + \text{c.c.}\,,
\end{equation}
%-------------------------------------------------
where the complex amplitude is defined as $\hat{\mathbf{F}}_{\text{j}}=|\hat{\mathbf{F}}_{\text{j}}|\mathrm{e}^{i \phi_{\text{j}}}$ with $j=\text{I},\text{O}$, and \emph{c.c.} denotes complex conjugate.

The average external work is given by
%-------------------------------------------------
\begin{equation}\label{Work}
    \overline{\mathsf{W}}=\frac{1}{4}\hat{\mathbf{F}}_{\text{O}}^{\dagger} \hat{\mathbf{X}}_{\text{O}} + \frac{1}{4}\hat{\mathbf{F}}_{\text{I}}^{\dagger} \hat{\mathbf{X}}_{\text{I}}+\text{c.c.}\,,
\end{equation}
%------------------------------------------------------
where the the operator $\dagger$ is the complex conjugate transpose.
Using the expression of $\hat{\mathbf{X}}_I$ in \eqref{eqXIc2}, 
%-------------------------------------------------
\begin{equation}\label{Work2}
    \overline{\mathsf{W}}=  \hat{\mathsf{W}}_0+\frac{1}{4}\hat{\mathbf{F}}_{\text{w}}^{\dagger} \hat{\mathbf{X}}_{\text{O}} +\text{c.c.}\,,\qquad \hat{\mathbf{F}}_{\text{w}}=\hat{\mathbf{F}}_{\text{O}}+\mathbf{K}_{\text{IO}}^{\dagger}\mathbf{A}^{\dagger}\mathbf{F}_{\text{I}}\,,\qquad \hat{\mathsf{W}}_0=-\frac{1}{4}\mathbf{F}_{\text{I}}^{\dagger}\mathbf{A}\hat{\mathbf{F}}_{\text{I}}\,.
\end{equation}
%------------------------------------------------------
The time average of the kinetic energy $\mathsf{K}$ can be written as
%------------------------------------------------------
\begin{equation}\label{kineticeff}
\overline{\mathsf{K}}=\hat{\mathsf{K}}_0+\frac{1}{8}\omega^2\hat{\mathbf{X}}_{\text{O}}^{\dagger}\mathbf{M}_{a}\hat{\mathbf{X}}_{\text{O}}-\frac{1}{4}\hat{\mathbf{F}}_{a}^{\dagger}\hat{\mathbf{X}}_{\text{O}}+\text{c.c.}\,,
\end{equation}
%-------------------------------------------------
where 
%-------------------------------------------------
\begin{equation}\label{Fa}
    \hat{\mathsf{K}}_0=\frac{\omega^2}{8}\hat{\mathbf{F}}_{\text{I}}^{\dagger}\mathbf{A}^{\dagger}\mathbf{M}_{\text{II}} \mathbf{A}\hat{\mathbf{F}}_{\text{I}}\,,\qquad
    \mathbf{M}_{a}=\mathbf{M}_{\text{OO}}+\mathbf{K}_{\text{IO}}^{\dagger}\mathbf{A}^{\dagger}\mathbf{M}_{\text{II}}\mathbf{A}\mathbf{K}_{\text{IO}},\qquad \mathbf{F}_{a}=\omega^2\mathbf{K}_{\text{IO}}^{\dagger} \mathbf{A}^{\dagger}\mathbf{M}_{\text{II}} \mathbf{A}\mathbf{F}_{\text{I}}\,.
\end{equation}
%-------------------------------------------------
Similarly, the time average of the potential energy  $\mathsf{P}$ reads
%-------------------------------------------------
\begin{equation}\label{Potentialeff2}
\overline{\mathsf{P}}=\hat{\mathsf{P}}_0+\frac{1}{8}\hat{\mathbf{X}}_{\text{O}}^{\dagger}\mathbf{K}_{b}\hat{\mathbf{X}}_{\text{O}}-\frac{1}{4}\hat{\mathbf{F}}_{b}^{\dagger}\hat{\mathbf{X}}_{\text{O}}+\text{c.c.}\,,
\end{equation}
%-------------------------------------------------
where the potential at rest is defined as
%-------------------------------------------------
\begin{equation}
    \hat{\mathsf{P}}_0=\frac{1}{8}\hat{\mathbf{F}}_{\text{I}}^{\dagger}\mathbf{A}^{\dagger}\mathbf{K}_{\text{II}} \mathbf{A}\hat{\mathbf{F}}_{\text{I}}\,.
\end{equation}
%-------------------------------------------------
Also
%-------------------------------------------------
\begin{equation}\label{Effstiffness}
\mathbf{K}_{b}=\mathbf{K}_{\text{OO}}+ \mathbf{K}_{\text{IO}}^{\dagger}\mathbf{A}^{\dagger}\mathbf{K}_{\text{II}} \mathbf{A}\mathbf{K}_{\text{IO}} +2 \mathbf{K}_{\text{IO}}^{\dagger}\mathbf{A}\mathbf{K}_{\text{IO}}\,,\qquad
\mathbf{F}_{b}=\mathbf{K}_{\text{IO}}^{\dagger}\mathbf{A}^{\dagger}\mathbf{K}_{\text{II}} \mathbf{A}\hat{\mathbf{F}}_{\text{I}} +\mathbf{K}_{\text{IO}}^{\dagger}\mathbf{A}\hat{\mathbf{F}}_{\text{I}}\,.
\end{equation}
%-------------------------------------------------
Thus, the time average of the Lagrangian in \eqref{Lagrang} reduces to
%-------------------------------------------------
\begin{equation}\label{Lagran2}
\begin{aligned}
    \overline{\mathsf{L}}=\overline{\mathsf{K}}-\overline{\mathsf{P}}-\overline{\mathsf{W}}
    =\frac{1}{8}\hat{\mathbf{X}}_{\text{O}}^{\dagger}\left(\omega^2\mathbf{M}_{a}-\mathbf{K}_{b}\right)\hat{\mathbf{X}}_{\text{O}}+\frac{1}{4}(-\hat{\mathbf{F}}_{a}+\hat{\mathbf{F}}_{b}+\hat{\mathbf{F}}_{\text{w}})^{\dagger}\hat{\mathbf{X}}_{\text{O}}+\hat{\mathsf{K}}_0-\hat{\mathsf{P}}_0-\hat{\mathsf{W}}_0+\text{c.c.}\,.
\end{aligned}
\end{equation}
%-------------------------------------------------
The time average of the Lagrangian of the effective-mass system in \eqref{effLagrang} is given by
%-------------------------------------------------
\begin{equation}\label{Lagranlump}
    \overline{\mathsf{L}}_{\text{eff}}=\omega^2\frac{1}{8}\hat{\mathbf{X}}_{\text{O}}^{\dagger}\mathbf{M}_{\text{eff}}\hat{\mathbf{X}}_{\text{O}}-\frac{1}{4}\hat{\mathbf{F}}_{\text{eff}}^{\dagger}\hat{\mathbf{X}}_{\text{O}}-\overline{\mathsf{P}}_{\text{eff}}\,.
\end{equation}
%-------------------------------------------------
Equating the two average Lagrangians~\eqref{Lagran2} and~\eqref{Lagranlump}, $\overline{\mathsf{L}}=\overline{\mathsf{L}}_{\text{eff}}$, yields
%-------------------------------------------------
\begin{equation}\label{Lagrangequivalence}
    \frac{1}{8}\hat{\mathbf{X}}_{\text{O}}^{\dagger}\left(\omega^2\mathbf{M}_{a}-\mathbf{K}_{b}\right)\hat{\mathbf{X}}_{\text{O}}+\frac{1}{4}(-\hat{\mathbf{F}}_{a}+\hat{\mathbf{F}}_{b}-\hat{\mathbf{F}}_{\text{w}})^{\dagger}\hat{\mathbf{X}}_{\text{O}}+\hat{\mathsf{K}}_0-\hat{\mathsf{P}}_0-\hat{\mathsf{W}}_0=\omega^2\frac{1}{8}\hat{\mathbf{X}}_{\text{O}}^{\dagger}\mathbf{M}_{\text{eff}}\hat{\mathbf{X}}_{\text{O}}-\frac{1}{4}\hat{\mathbf{F}}_{\text{eff}}^{\dagger}\hat{\mathbf{X}}_{\text{O}}-\overline{\mathsf{P}}_{\text{eff}}\,.
\end{equation}
%-------------------------------------------------
This implies that
%-------------------------------------------------
\begin{equation}\label{effmass}
	\mathbf{M}_{\text{eff}}=\mathbf{M}_{\text{a}}-\frac{\mathbf{K}_{\text{b}}}{\omega^2},\qquad 
	\hat{\mathbf{F}}_{\text{eff}}=\hat{\mathbf{F}}_{\text{w}}+\hat{\mathbf{F}}_{\text{a}}-\hat{\mathbf{F}}_{\text{b}}\,,\qquad
	\mathsf{P}_{\text{eff}}=\hat{\mathsf{P}}_0+\hat{\mathsf{W}}_0-\hat{\mathsf{K}}_0
	=\frac{3}{2}\hat{\mathsf{W}}_0
	=-\frac{3}{8}\hat{\mathbf{F}}_{\text{I}}^{\dagger}\mathbf{A}\hat{\mathbf{F}}_{\text{I}}\,.
\end{equation}
%-------------------------------------------------
Here, $\mathsf{P}_{\text{eff}}$ is interpreted as potential energy at rest due to inner forces when there is no outer motion. Such a potential vanishes in the absence of inner forces, i.e., when $\hat{\mathbf{F}}_{\text{I}}=\mathbf{0}$. 

Using \eqref{Effstiffness}$_1$, the effective mass matrix in \eqref{effmass}$_1$ can be simplified to read 
%-------------------------------------------------
\begin{equation}
\begin{aligned}
	\mathbf{M}_{\text{eff}}&=\mathbf{M}_{\text{a}}-\frac{\mathbf{K}_{\text{b}}}{\omega^2}\\
	&=\mathbf{M}_{\text{OO}}+\mathbf{K}_{\text{IO}}^{\dagger}\mathbf{A}^{\dagger}\mathbf{M}_{\text{II}}
	\mathbf{A}\mathbf{K}_{\text{IO}}-\frac{1}{\omega^2}\mathbf{K}_{\text{IO}}^{\dagger}\mathbf{A}^{\dagger}
	\mathbf{K}_{\text{II}} \mathbf{A}\mathbf{K}_{\text{IO}} -\frac{2}{\omega^2} \mathbf{K}_{\text{IO}}
	\mathbf{A}\mathbf{K}_{\text{IO}}-\frac{1}{\omega^2}\mathbf{K}_{\text{OO}}\,\\
 &=\mathbf{M}_{\text{OO}}+\frac{1}{\omega^2}\mathbf{K}_{\text{IO}}^{\dagger}
	\mathbf{A}^{\dagger}\left(\omega^2\mathbf{M}_{\text{II}}-\mathbf{K}_{\text{II}}\right)\mathbf{A}
	\mathbf{K}_{\text{IO}} -\frac{2}{\omega^2} \mathbf{K}_{\text{IO}}^{\dagger}\mathbf{A}
	\mathbf{K}_{\text{IO}}-\frac{1}{\omega^2}\mathbf{K}_{\text{OO}}\,\\
 &=\mathbf{M}_{\text{OO}}+\frac{1}{\omega^2}\mathbf{K}_{\text{IO}}^{\dagger}
	\mathbf{A}^{\dagger}\mathbf{A}^{\rm -T}\mathbf{A}\mathbf{K}_{\text{IO}} -\frac{2}{\omega^2} 
	\mathbf{K}_{\text{IO}}^{\dagger}\mathbf{A}\mathbf{K}_{\text{IO}}-\frac{1}{\omega^2}\mathbf{K}_{\text{OO}}\\
 &=\mathbf{M}_{\text{OO}}+\frac{1}{\omega^2}\mathbf{K}_{\text{IO}}^{\dagger}
	\mathbf{A}\mathbf{K}_{\text{IO}} -\frac{2}{\omega^2} \mathbf{K}_{\text{IO}}^{\dagger}
	\mathbf{A}\mathbf{K}_{\text{IO}}-\frac{1}{\omega^2}\mathbf{K}_{\text{OO}}\\
 &=\mathbf{M}_{\text{OO}}
	-\frac{1}{\omega^2}\mathbf{K}_{\text{IO}}^{\dagger}\mathbf{A}\mathbf{K}_{\text{IO}}
	-\frac{1}{\omega^2}\mathbf{K}_{\text{OO}} \\
 & = \mathbf{M}_{\text{OO}} - \frac{1}{\omega^2}\left[\mathbf{K}_{\text{OI}}\left(\omega^2\mathbf{M}_{\text{II}} - \mathbf{K}_{\text{II}} \right)^{-1}\mathbf{K}_{\text{IO}} + \mathbf{K}_{\text{OO}}\right]\,, \label{Eq:Meff:Chiral:AE}
 \end{aligned}
\end{equation}
%------------------------------------------------
where use has been made of the relations: $\mathbf{K}_{\text{IO}}^{\dagger}=\mathbf{K}_{\text{IO}}^{\mathsf{T}}=\mathbf{K}_{\text{OI}}$, and $\mathbf{A}^{-1}=\left(\omega^2\mathbf{M}_{\text{II}} - \mathbf{K}_{\text{II}} \right)$. The effective mass matrix above is identical to that derived from the momentum equivalence in \eqref{massFourier}.  

Using \eqref{Work2}$_2$,~\eqref{Fa}$_3$,~\eqref{Effstiffness}$_2$, the effective force vector  in \eqref{Effstiffness}$_2$ can be simplified to read
%-------------------------------------------------
\begin{equation}\label{effforce2}
\begin{aligned}
 	\mathbf{F}_{\text{eff}} &=\hat{\mathbf{F}}_{\text{O}} + \mathbf{K}_{\text{IO}}^{\dagger} 
	\mathbf{A}^{\dagger}(\omega^2\mathbf{M}_{\text{II}}-\mathbf{K}_{\text{II}}) \mathbf{A}\mathbf{F}_{\text{I}}\\
	&= \hat{\mathbf{F}}_{\text{O}} + \mathbf{K}_{\text{IO}}^{\dagger} \mathbf{A}^{\dagger}
	\mathbf{A}^{-1} \mathbf{A}\mathbf{F}_{\text{I}}\\
	&= \hat{\mathbf{F}}_{\text{O}} + \mathbf{K}_{\text{IO}}^{\dagger} \mathbf{A}^{\dagger}\mathbf{F}_{\text{I}}\\
	&= \hat{\mathbf{F}}_{\text{O}} + \mathbf{K}_{\text{OI}}\left(\omega^2\mathbf{M}_{\text{II}} 
	- \mathbf{K}_{\text{II}} \right)^{-1}\mathbf{F}_{\text{I}}  \,,
\end{aligned}
\end{equation}
%-------------------------------------------------
%-----------------------------
where we have used the relations: $\mathbf{A}^{\dagger}=\mathbf{A}^{\mathsf{T}}=\mathbf{A}=\left(\omega^2\mathbf{M}_{\text{II}} - \mathbf{K}_{\text{II}} \right)^{-1}$, and $\mathbf{K}_{\text{IO}}^{\dagger}=\mathbf{K}_{\text{IO}}^{\mathsf{T}}=\mathbf{K}_{\text{OI}}$. The effective force vector above is also the same as that derived from the momentum equivalence in \eqref{forcingFourier}.

%------------------------------------------------------
%------------------------------------------------------
\subsubsection{Dynamic condensation}

Assume periodic forcing with given frequency $\omega$, that is $\mathbf{F}_{\text{O}}(t)=\hat{\mathbf{F}}_{\text{O}}(\omega)\,\mathrm{e}^{i\omega t}$, and $\mathbf{F}_{\text{I}}(t)=\hat{\mathbf{F}}_{\text{I}}(\omega)\,\mathrm{e}^{i\omega t}$.
Then, $\mathbf{X}_{\text{O}}(t)=\hat{\mathbf{X}}_{\text{O}}(\omega)\,\mathrm{e}^{i\omega t}$, and $\mathbf{X}_{\text{I}}(t)=\hat{\mathbf{X}}_{\text{I}}(\omega)\,\mathrm{e}^{i\omega t}$, where $\hat{\mathbf{X}}(\omega)$ is the complex Fourier amplitude of $\mathbf{X}$. 
Fourier transforming the dynamical equations of the outer and inner systems in \eqref{maeq} yields
%------------------------------------------------------
\begin{equation}\label{eqXO}
\begin{aligned}
     -\omega^2\mathbf{M}_{\text{OO}}\hat{\mathbf{X}}_{\text{O}} + \mathbf{K}_{\text{OO}}\hat{\mathbf{X}}_{\text{O}} +\mathbf{K}_{\text{OI}}\hat{\mathbf{X}}_{\text{I}} &=\hat{\mathbf{F}}_{\text{O}}\,, \\
     -\omega^2\mathbf{M}_{\text{II}}\hat{\mathbf{X}}_{\text{I}} + \mathbf{K}_{\text{IO}}\hat{\mathbf{X}}_{\text{O}} +\mathbf{K}_{\text{II}}\hat{\mathbf{X}}_{\text{I}}&=\hat{\mathbf{F}}_{\text{I}}\,.
\end{aligned}
\end{equation}
In this approach the inner (micro) variables are eliminated as follows.
From \eqref{eqXO}, we solve for $\hat{\mathbf{X}}_{\text{I}}$:
%------------------------------------------------------
\begin{equation}\label{eqXIc}
    \hat{\mathbf{X}}_{\text{I}}=\left(\omega^2\mathbf{M}_{\text{II}} - \mathbf{K}_{\text{II}} \right)^{-1}(\mathbf{K}_{\text{IO}}\hat{\mathbf{X}}_{\text{O}}-\hat{\mathbf{F}}_{\text{I}})\,.
\end{equation}
%------------------------------------------------------
This can be rewritten as 
%------------------------------------------------------
\begin{equation}\label{eqXIc2}
    \hat{\mathbf{X}}_{\text{I}}=\mathbf{A}\mathbf{K}_{\text{IO}}\hat{\mathbf{X}}_{\text{O}} - \mathbf{A}\hat{\mathbf{F}}_{\text{I}}\,,
    \qquad \mathbf{A}=\left(\omega^2\mathbf{M}_{\text{II}} - \mathbf{K}_{\text{II}} \right)^{-1}\,.
\end{equation}
%------------------------------------------------------
Plugging \eqref{eqXIc2} into \eqref{eqXO} yields
%------------------------------------------------------
\begin{equation}
-\omega^2\mathbf{M}_{\text{OO}}\hat{\mathbf{X}}_{\text{O}} + \mathbf{K}_{\text{OO}}\hat{\mathbf{X}}_{\text{O}} +\mathbf{K}_{\text{OI}}\mathbf{A}\mathbf{K}_{\text{IO}}\hat{\mathbf{X}}_{\text{O}} - \mathbf{K}_{\text{OI}}\mathbf{A}\hat{\mathbf{F}}_{\text{I}}=\hat{\mathbf{F}}_{\text{O}}\,,
\end{equation}
%------------------------------------------------------
which can be rewritten as
%------------------------------------------------------
\begin{equation}\label{eqXOb}
    -\omega^2\left(\mathbf{M}_{\text{OO}} - \frac{1}{\omega^2}\mathbf{K}_{\text{OI}}\mathbf{A}\mathbf{K}_{\text{IO}}  \right) \hat{\mathbf{X}}_{\text{O}} + \mathbf{K}_{\text{OO}}\hat{\mathbf{X}}_{\text{O}}=\hat{\mathbf{F}}_{\text{O}} +  \mathbf{K}_{\text{OI}}\mathbf{A}\hat{\mathbf{F}}_{\text{I}}\,.
\end{equation}
%-------------------------------------------------
Therefore, one can lump the inertial forces of the inner system into those of the outer system and define the condensed macro system
%------------------------------------------------------
\begin{equation}\label{eqXOb}
-\omega^2\mathbf{M}_{\text{eff}}(\omega)\hat{\mathbf{X}}_O=\hat{\mathbf{F}}_{\text{eff}}(\omega)\,,
\end{equation}
%-------------------------------------------------
where the effective mass matrix is written as
%------------------------------------------------------
\begin{equation} \label{Eq:Meff:DC}
 \mathbf{M}_{\text{eff}}(\omega)  =\mathbf{M}_{\text{OO}} - \frac{1}{\omega^2}\left[\mathbf{K}_{\text{OI}}\left(\omega^2\mathbf{M}_{\text{II}} - \mathbf{K}_{\text{II}} \right)^{-1}\mathbf{K}_{\text{IO}} + \mathbf{K}_{\text{OO}}\right]\,, 
\end{equation}
%------------------------------------------------------
and the effective force vector reads
%------------------------------------------------------
\begin{equation} \label{Eq:Feff:DC}
     \hat{\mathbf{F}}_{\text{eff}}(\omega)  
     = \hat{\mathbf{F}}_{\text{O}} + \mathbf{K}_{\text{OI}}\left(\omega^2\mathbf{M}_{\text{II}} 
	- \mathbf{K}_{\text{II}} \right)^{-1}\mathbf{F}_{\text{I}} \,.
\end{equation}
%-------------------------------------------------

\begin{remark}[]
The effective mass and force vector obtained using the three different approaches, namely, momentum equivalence (Eqs.~\eqref{massFourier} and \eqref{forcingFourier}), action equivalence (Eqs.~\eqref{Eq:Meff:Chiral:AE} and \eqref{effforce2}), and dynamic condensation (Eqs.~\eqref{Eq:Meff:DC} and \eqref{Eq:Feff:DC}), are identical. In particular, both the effective mass matrix and force vector are frequency-dependent and include an added mass matrix and an added force vector that account for the momentum of the microstructure. 
\end{remark}

\paragraph{Negative effective mass}\label{Sec:negativemass}
The general effective mass matrix given in \eqref{massFourier} may have negative eigenvalues for certain frequencies. In our case, the eigen-masses are real. The natural frequencies $\omega_i$ of the inner, or microstructure, satisfy the eigenvalue problem $|\omega^2\mathbf{M}_{\text{II}} - \mathbf{K}_{\text{II}}|=0$.
Note that the outer mass and stiffness matrices $\mathbf{M}_{\text{OO}}$ and $\mathbf{K}_{\text{OO}}$ are positive-definite. If the matrix $\mathbf{K}_{\text{OI}}\left(\omega^2\mathbf{M}_{\text{II}} - \mathbf{K}_{\text{II}} \right)^{-1}\mathbf{K}_{\text{IO}}$ is positive-definite, then the effective mass may have negative eigenvalues. Positive-definiteness of $\mathbf{K}_{\text{OI}}\left(\omega^2\mathbf{M}_{\text{II}}-\mathbf{K}_{\text{II}} \right)^{-1}\mathbf{K}_{\text{IO}}$ is equivalent to $\mathbf{w}^{\dagger}\left(\omega^2\mathbf{M}_{\text{II}}-\mathbf{K}_{\text{II}}\right)^{-1}
\mathbf{w}>0$, for $\mathbf{w}=\mathbf{K}_{\text{IO}}\mathbf{v}$, and arbitrary $\mathbf{v}$.
Therefore, a necessary condition for the effective mass matrix to have negative eigenvalues is $\omega^2\mathbf{M}_{\text{II}} - \mathbf{K}_{\text{II}}$ being positive-definite.
This happens when the excitation frequency $\omega$ is greater than all of the natural frequencies $\omega_i$ of the microstructure, i.e., $\omega>\mathsf{max}\{(\omega_i)_1,\hdots (\omega_i)_{N_I}\}$, where $N_I$ is the number of the inner degrees of freedom.
Thus, one or more eigen-masses of the effective mass matrix may become negative when the excitation frequency $\omega$ approaches one of the natural frequencies $\omega_i$ from above. Clearly, if the frequency ranges overlap, the condition that all the eigenvalues of the mass matrix are negative is met.

Sufficient conditions for at least one negative eigen-mass follow from  the Gershgorin circle theorem \citep{Horn2012}. Given the effective mass matrix $\mathbf{M}_{\text{eff}}=[M_{ij}]$, an eigen-mass $\lambda$ lies within the closed discs of the complex plane $(\mathrm{Re}\lambda,\mathrm{Im}\lambda)$
\begin{equation}
\big|\lambda - M_{ii}\big|\le \sum_{i\ne j} \big| M_{ij}\big|=R_i\,\quad i=1,\cdots N_I\,,\label{Gershgorin}
\end{equation}
centered at $M_{ii}$ with radius $R_{i}$. Thus, sufficient conditions to have one negative eigen-mass is when one of the two Gershgorin discs lies in the negative part of the complex plane ($\mathrm{Re}\lambda<0$). 

\paragraph{Effective stiffness matrix}
The two terms $\mathbf{M}_{\text{a}}$ and $\mathbf{K}_{\text{b}}/\omega^2$ of the effective mass in \eqref{effmass}$_1$ suggest another formalism that defines $\mathbf{K}_{\text{b}}$ as an effective stiffness matrix  and $\mathbf{M}_{\text{a}}$ as an alternative effective mass matrix. In particular, consider an equivalent lumped mass-spring system with effective mass $\mathbf{\widetilde{M}}_{\text{eff}}$, stiffness matrix $\mathbf{K}_{\text{eff}}$, potential energy at rest $\mathsf{P}_{\text{eff}}$ subject to the effective force $\mathbf{F}_{\text{eff}}$. The associated Lagrangian is given by
%------------------------------------------------------
\begin{equation}\label{effLagrang10}
 	\mathsf{L}_{\text{eff}}=\frac{1}{2}\dot{\mathbf{X}}_{\text{O}}^{\mathsf{T}}\mathbf{\widetilde{M}}_{\text{eff}}
\dot{\mathbf{X}}_{\text{O}}-\frac{1}{2}\mathbf{X}_{\text{O}}^{\mathsf{T}}\mathbf{K}_{\text{eff}}
\mathbf{X}_{\text{O}} - \mathsf{P}_{\text{eff}} - \mathbf{F}_{\text{eff}}^{\mathsf{T}}\mathbf{X}_{\text{O}}\,.
\end{equation}
%-------------------------------------------------
The time average of $\mathsf{L}_{\text{eff}}$ follows as
%-------------------------------------------------
\begin{equation}\label{Lagranlump4}
\overline{\mathsf{L}}_{\text{eff}}=\omega^2\frac{1}{8}\hat{\mathbf{X}}_{\text{O}}^{\dagger}\mathbf{\widetilde{M}}_{\text{eff}}\hat{\mathbf{X}}_{\text{O}}-\frac{1}{8}\hat{\mathbf{X}}_{\text{O}}^{\dagger}\mathbf{K}_{\text{eff}}\hat{\mathbf{X}}_{\text{O}}-\overline{\mathsf{P}}_{\text{eff}}-\frac{1}{4}\hat{\mathbf{F}}_{\text{eff}}^{\dagger}\hat{\mathbf{X}}_{\text{O}}+\text{c.c.}\,.
\end{equation}
%-------------------------------------------------
Equating the two averaged Lagrangians~\eqref{Lagran2} and~\eqref{Lagranlump4} yields the same effective force and potential energy at rest as in \eqref{effmass}$_{2,3}$, but a different effective mass given by \eqref{Fa}$_2$, that is
%-------------------------------------------------
\begin{equation}\label{effmass4}
\mathbf{\widetilde{M}}_{\text{eff}}(\omega)=\mathbf{M}_{\text{a}}=\mathbf{M}_{\text{OO}}+\mathbf{K}_{\text{IO}}^{\dagger}\mathbf{A}^{\dagger}\mathbf{M}_{\text{II}}\mathbf{A}\mathbf{K}_{\text{IO}}\,.
\end{equation}
%-------------------------------------------------
The effective stiffness matrix follows from \eqref{Effstiffness}$_1$ as
\begin{equation}\label{effmass6}
 \mathbf{K}_{\text{eff}}(\omega)=\mathbf{K}_{\text{b}}=\mathbf{K}_{\text{OO}}+ \mathbf{K}_{\text{IO}}^{\mathsf{T}}(\mathbf{A}\mathbf{K}_{\text{II}} \mathbf{A}+2\mathbf{A})\mathbf{K}_{\text{IO}}\,.
\end{equation}
%-------------------------------------------------
The effective mass of the equivalent lumped-mass system in \eqref{effmass}$_1$  can be written as
\begin{equation}
 \mathbf{M}_{\text{eff}}(\omega)=\mathbf{M}_{\text{a}}-\frac{\mathbf{K}_{\text{b}}}{\omega^2}= \mathbf{\widetilde{M}}_{\text{eff}}(\omega)-\frac{\mathbf{K}_{\text{eff}}(\omega)}{\omega^2}\,.
\end{equation}
%-------------------------------------------------

%-----------------------------
%-----------------------------
\begin{figure}[t!]
\centering
\includegraphics[width=0.89\textwidth]{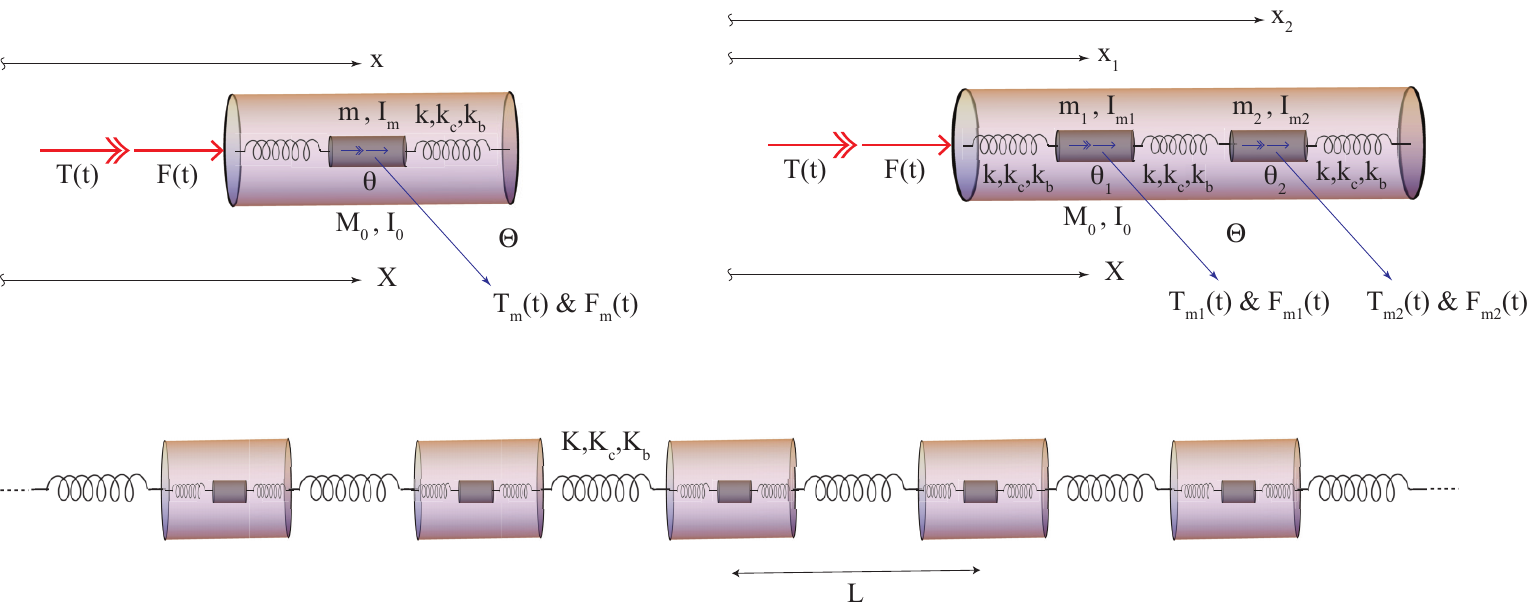}
\vspace*{0.0in}
\caption{Bottom panel: A $1$D composite lattice with microstructure.  Cells are separated by distance $L$ and connected to each other by outer chiral springs with elastic constants $K$, $K_c$, and $K_t$. Top panel left: The microstructure of the lattice cell is a $1$D chiral spring-interconnected mass-in-mass system. The macro element has mass $M_0$ and mass moment of inertial $I_0$. There is a single micro element inside with mass $m$ and mass moment of inertial $I_m$ that is connected to the macro element by two identical linear chiral springs with elastic constants $k$, $k_c$, and $k_t$. Top panel right: A macro element with mass $M_0$ and mass moment of inertial $I_0$. There are two micro elements inside each with mass $m$ and mass moment of inertial $I_m$ that are connected to the macro element and each other by three identical linear chiral springs.
}
\label{Unit-Cell2}
\end{figure}
%-----------------------------
%-----------------------------

%-----------------------------
%-----------------------------
\section{$1$D composite lattices with $1$D chiral spring-interconnected mass-in-mass microstructure}\label{Sec:1Dcompositelattice}

Consider the $1$D composite lattice made of $N$ identical cells depicted in the bottom panel of Fig.~\ref{Unit-Cell2}. The microstructure is made of a single mass-in-mass cell shown in the top-left panel of the same figure~(single micro-mass). The cells are separated by distance $L$ and are connected to each other by outer chiral springs with elastic constants $K$, $K_c$, and $K_t$. 
The single micro-mass cell is a hollow circular shaft with mass $M_0$ and mass moment of inertia $I_0$. Inside the unit cell there is a solid cylinder with mass $m$ and mass moment of inertia $I_m$ as depicted in the top-left panel of Fig.~\ref{Unit-Cell2}. This solid cylinder is connected to the hollow shaft by two identical chiral linear (micro) springs with the elastic constants $k$, $k_c$, and $k_t$.\footnote{See \citet{Yu2023} for an example of a $3$D structure that can be approximated by this chiral model.} Denoting the change in length of the spring by $\delta$ and its twist by $\theta$ the force and torque in the spring are written as
%---------------------
\begin{equation} 
	f=k\,\delta+k_c\,\theta\,,\qquad t=k_c\,\delta+k_t\,\theta\,.
\end{equation}
%---------------------  

In what follows, we will first derive the dynamical equations of the single mass-in-mass cell microstructure and derive its effective mass matrix, and in so doing, identify frequency ranges of negative mass. In \S\ref{Sec:two_micro_masses} we will consider a microstructure with two micro-masses depicted in the top-right panel of Fig.~\ref{Unit-Cell2}. We will show that the associated effective mass matrix is equivalent to that of some single micro-mass cell. This result can be generalized to a unit cell with $N$ micro-masses in series. This implies that in our study it suffices to consider only the single mass-in-mass unit cell depicted in the top-left panel of Fig.~\ref{Unit-Cell2}. Then, in \S\ref{Sec:Blochanalysis} we will explore the spectral properties~(frequency bands) of the $1$D composite lattice by way of a Bloch analysis. In \S\ref{Sec:defects} localized modes due to defects will be investigated.

%-----------------------------
%-----------------------------
\subsection{Effective mass matrix of a cell with a single micro-mass} \label{Sec:Lattice}

We model the microstructure of the cell with a single micro-mass as the $1$D chiral spring interconnected mass-in-mass system depicted in top-left panel of Fig.~\ref{Unit-Cell2}. The macro and micro generalized coordinates are $(X(t),\Theta(t))$ and  $(x(t),\theta(t))$, respectively. The balance of micro linear and angular momenta read
%---------------------
\begin{equation}  \label{Micro-Momenta}
\begin{aligned}
	-2f(t)+F_m(t)=-2k\big(x(t)-X(t)\big)-2k_c\big(\theta(t)-\Theta(t)\big)+F_m(t) &=m\,\ddot{x}(t)\,,\\
	-2t(t)+T_m(t)=-2k_c\big(x(t)-X(t)\big)-2k_t\big(\theta(t)-\Theta(t)\big)+T_m(t) &=I_m\,\ddot{\theta}(t) \,,
\end{aligned}
\end{equation}
%---------------------  
where $F_m(t)$ and $T_m(t)$ are the applied micro-force and micro-moment, respectively. Similarly, the balance of macro linear and angular momenta read
%---------------------
\begin{equation}  \label{Macro-Momenta}
\begin{aligned}
	F(t)+2f(t) = F(t)+2k\big(x(t)-X(t)\big)+2k_c\big(\theta(t)-\Theta(t)\big) &=M_0\,\ddot{X}(t)\,,\\
	T(t)+2t(t) = T(t)+2k_c\big(x(t)-X(t)\big)+2k_t\big(\theta(t)-\Theta(t)\big) &= I_0\,\ddot{\Theta}(t) \,,
\end{aligned}
\end{equation}
%--------------------- 
where $F(t)$ and $T(t)$ are the applied macro-force and macro-moment, respectively.
The dynamical equations~\eqref{Micro-Momenta}~and~\eqref{Macro-Momenta} can be written in the compact matrix form of \eqref{IOmatrixsys}. In particular, let us define the inner and outer generalized displacements 
%------------------------------------------------------
\begin{equation}\label{gendisp}
    \mathbf{X}_{\text{I}}=
    \begin{bmatrix}
   x(t)\\
    \theta(t)
    \end{bmatrix}\,,\qquad
    \mathbf{X}_{\text{O}}=
    \begin{bmatrix}
   X(t)\\
    \Theta(t)
    \end{bmatrix}\,,
\end{equation}
%------------------------------------------------------
and the inner and outer mass matrices
%------------------------------------------------------
\begin{equation}\label{massmatrices}
    \mathbf{M}_{\text{II}}=
    \begin{bmatrix}
   m & 0\\
    0 & I_m
    \end{bmatrix}\,,\qquad
    \mathbf{M}_{\text{OO}}=
    \begin{bmatrix}
   M_0 & 0\\
    0 & I_0
    \end{bmatrix}\,.
\end{equation}
%------------------------------------------------------
The stiffness matrices are written as $\mathbf{K}_{\text{OO}}=\mathbf{K}_{\text{II}}=2\mathbf{K}$, and $\mathbf{K}_{\text{OI}}=\mathbf{K}_{\text{IO}}^{\mathsf{T}}=-2\mathbf{K}$, where 
%------------------------------------------------------
\begin{equation}\label{stiffmatrixK}
\mathbf{K}=
    \begin{bmatrix}
   k & k_c\\
    k_c & k_t
    \end{bmatrix}\,,
\end{equation}
%------------------------------------------------------
and the outer and inner force vectors are defined as
%------------------------------------------------------
\begin{equation}\label{forcingA}
\mathbf{F}_{\text{O}}=
    \begin{bmatrix}
   F(t) \\
    T(t) 
    \end{bmatrix}\,,\qquad\mathbf{F}_{\text{I}}=
    \begin{bmatrix}
    F_m \\
    T_m
    \end{bmatrix}\,.
\end{equation}
%------------------------------------------------------

%------------------------------------------------------
%------------------------------------------------------
\subsubsection{The effective mass matrix and force vector}

We now assume that the macro and micro forces and torques are harmonic, i.e., $F(t)=\hat{F}\,e^{i\,\omega\,t}$, $T(t)=\hat{T}\,e^{i\,\omega\,t}$, $F_m(t)=\hat{F}_m\,e^{i\,\omega\,t}$, and $T_m(t)=\hat{T}_m\,e^{i\,\omega\,t}$. This implies that $X(t)=\hat{X}\,e^{i\,\omega\,t}$, $\Theta(t)=\hat{\Theta}\,e^{i\,\omega\,t}$, $x(t)=\hat{x}\,e^{i\,\omega\,t}$, and $\theta(t)=\hat{\theta}\,e^{i\,\omega\,t}$.
%------------------------------------------------------
Then, the Fourier transform of the generalized displacements vectors read
%------------------------------------------------------
\begin{equation}\label{Fouriergendisp}
\hat{\mathbf{X}}_{\text{O}}=
    \begin{bmatrix}
   \hat{X}\\
    \hat{\Theta}
    \end{bmatrix}\,,\qquad\hat{\mathbf{X}}_{\text{I}}=
    \begin{bmatrix}
   \hat{x}\\
    \hat{\theta}
    \end{bmatrix}\,.
\end{equation}
%------------------------------------------------------
The effective mass follows from \eqref{massFourier} as
%------------------------------------------------------
\begin{equation}
\begin{aligned}
    \mathbf{M}_{\text{eff}}(\omega) &=\mathbf{M}_{\text{OO}} - \frac{1}{\omega^2}\left[\mathbf{K}_{\text{OI}}
    \left(\omega^2\mathbf{M}_{\text{II}} - \mathbf{K}_{\text{II}} \right)^{-1}\mathbf{K}_{\text{IO}} 
    + \mathbf{K}_{\text{OO}}\right]\\
    &=\mathbf{M}_{\text{OO}} - \frac{2}{\omega^2}\mathbf{K}\left[\left(\omega^2\mathbf{M}_{\text{II}} 
    - 2\mathbf{K} \right)^{-1}2\mathbf{K} + \mathbf{I}\right]\,,
\end{aligned}
\end{equation}
%-------------------------------------------------
where $\mathbf{I}$ is the $2\times 2$~identity matrix. Thus 
%-------------------------------------------------
\begin{equation}\label{1mass_effmass}
	\mathbf{M}_{\text{eff}}(\omega)= \begin{bmatrix}
	M_0 +\frac{2 m(2 k_c^2 -2 k k_t + I_m k \omega^2)}{4 k_c^2 - (2 k_t - I_m \omega^2) (2 k - m \omega^2)} 
	& \frac{ 2 I_m k_c m \omega^2 }{4 k_c^2 - (2 k_t - I_m \omega^2) (2 k - m \omega^2)} \\
	\frac{ 2 I_m k_c m \omega^2 }{4 k_c^2 - (2 k_t - I_m \omega^2) (2 k - m \omega^2)} 
	& I_0 +\frac{2 I_m( 2 k_c^2 -2 k k_t + m k_t \omega^2)}{4 k_c^2 - (2 k_t - I_m \omega^2) (2 k - m \omega^2)}
    \end{bmatrix}\,.
\end{equation}
%-------------------------------------------------
Note that the effective mass matrix can be written as a static mass matrix plus an added mass matrix, i.e., 
\begin{equation}
   \mathbf{M}_{\text{eff}}(\omega)=\mathbf{M}_{\text{static}}+\mathbf{M}_{\text{added}}(\omega)\,,  
\end{equation}
%-------------------------------------------------
where
%-------------------------------------------------
\begin{equation} 
	\mathbf{M}_{\text{static}}=\begin{bmatrix}
	M_0 & 0 \\
	0 & I_0
	\end{bmatrix}
	\,,\qquad
	\mathbf{M}_{\text{added}}(\omega)=\begin{bmatrix}
	M(\omega) & J(\omega) \\
	J(\omega) & I(\omega)
	\end{bmatrix}
	\,,
\end{equation}
%--------------------- 
and 
%---------------------
\begin{equation}\label{Mcoeff2}  
\begin{aligned}
	M(\omega) &= \frac{2 m \left(I_m \,k\, \omega ^2-2 k \,k_t+2 k_c^2\right)}
	{4k_c^2-\left(2 k_t-I_m \,\omega^2\right) \left(2 k-m \omega^2\right)}   \,,\\
	J(\omega) &=  \frac{2m\, I_m k_c \, \omega^2}{4 k_c^2-\left(2 k_t-I_m \omega^2\right) 
	\left(2 k-m \omega^2\right)}  \,,\\
	I(\omega) &= \frac{2I_m \left(-2 k\,k_t+2k_c^2+k_t\, m\, \omega^2\right)}
	{4k_c^2-\left(2k_t-I_m\, \omega^2\right) \left(2k-m\, \omega ^2\right)}  \,.
\end{aligned}
\end{equation}
%---------------------  
From\eqref{forcingFourier}, the effective force vector reads 
%------------------------------------------------------
\begin{equation} \label{forcingC}
	\mathbf{F}_{\text{eff}}(\omega) =\hat{\mathbf{F}}_{\text{O}} +\mathbf{K}_{\text{OI}}
	\left(\omega^2\mathbf{M}_{\text{II}} - \mathbf{K}_{\text{II}} \right)^{-1}\hat{\mathbf{F}}_{\text{I}}    \,,
\end{equation}
%-----------------------------------------------------
where
%------------------------------------------------------
\begin{equation}\label{forcingC}
\hat{\mathbf{F}}_{\text{O}}(\omega)=
    \begin{bmatrix}
   \hat{F} \\
    \hat{T} 
    \end{bmatrix}\,,\qquad\hat{\mathbf{F}}_{\text{I}}(\omega)=
    \begin{bmatrix}
    \hat{F}_m \\
    \hat{T}_m
    \end{bmatrix}\,,
\end{equation}
%------------------------------------------------------
are the Fourier transforms of the micro and macro force vectors.

%-------------------------------------------------
\subsubsection{Frequency ranges of negative effective mass}

\begin{figure}[t!]
\centering
\includegraphics[width=0.65\textwidth]{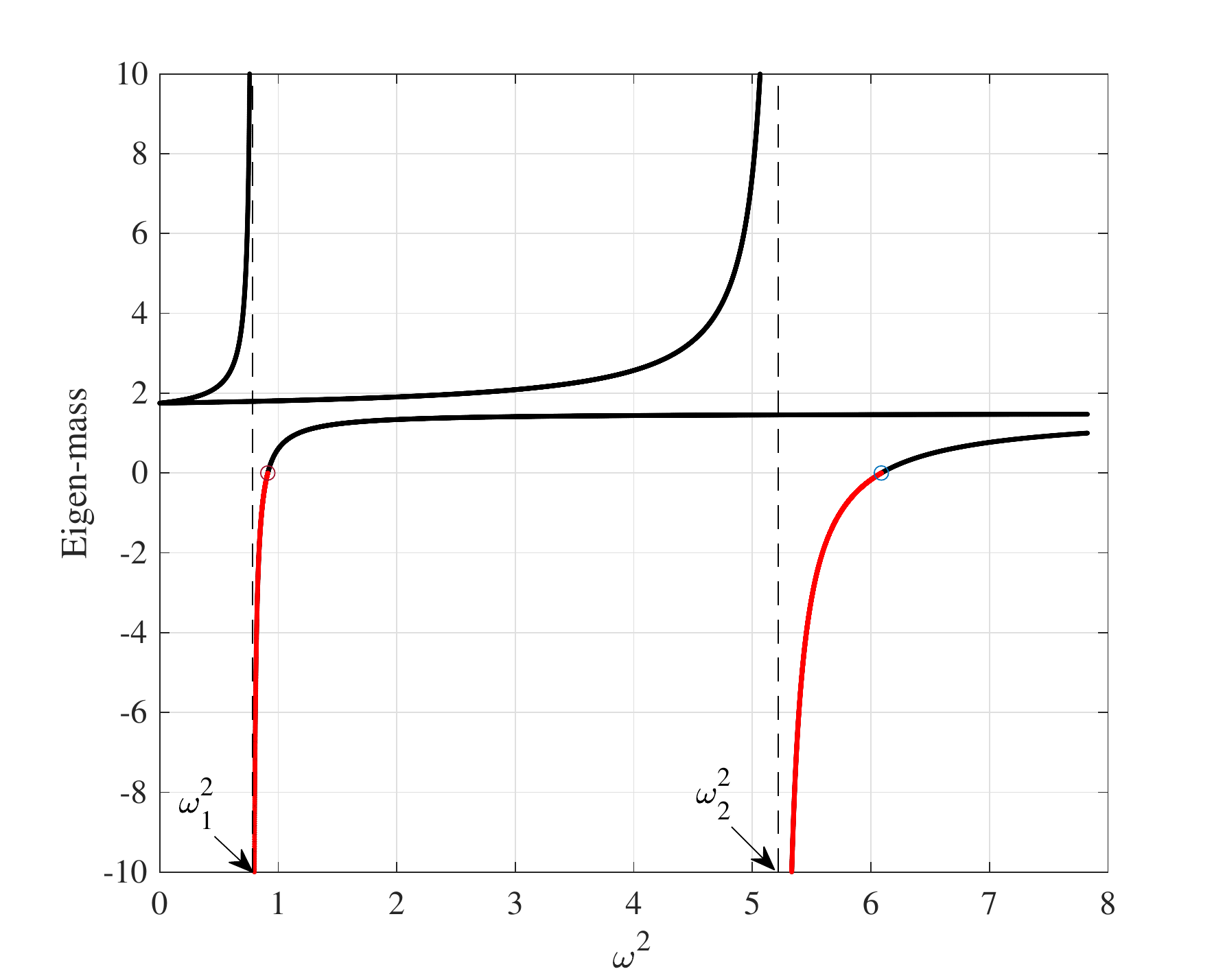}
\vspace*{-0.0in}
\caption{Frequency ranges (red) of negative eigen-masses occur for excitation frequencies~$\omega$ approaching from above the natural frequencies $\omega_i$ of the microstructure with effective mass matrix given in \eqref{1mass_effmass}. The circles indicate where the eigen-mass is zero. Inner paramerers $m=0.25,I_m=0.25,k=1/2,k_t=1/4$, and outer parameters $M_0=1.5,I_0=1.5, K=2,K_t=2$ are chosen. The inner and outer chiral stiffnesses are selected to avoid zero-energy modes, $k_c=\gamma\sqrt{k k_t}$, and $K_c=\Gamma\sqrt{K K_t}$ with $\gamma=\Gamma=0.7$.} 
\label{negativeeigenmasses}
\end{figure}

In \S\ref{Sec:negativemass} we  showed that an effective eigen-mass~$\lambda$ of $\mathbf{M}_{\text{eff}}$ is negative when the excitation frequency $\omega$ is greater than any of the natural frequencies $\omega_i$ of the microstructure, where $|\omega_i^2\mathbf{M}_{\text{II}} - \mathbf{K}_{\text{II}}|=0$.
For the specific microstructure considered above, the $2\times 2$ effective mass matrix  in \eqref{1mass_effmass} has a negative eigen-mass in the range $\omega_i<\omega<(1+\alpha)\omega_i$, for each natural inner frequency~($i=1,2$). Here, $(1+\alpha)\omega_i$ is greater than $\omega_i$~($\alpha>0$), where the effective eigen-mass vanishes, that is $\lambda\big((1+\alpha)\omega_i\big)=0$. As an example, in Fig.~\ref{negativeeigenmasses} we depict the two eigen-masses of the microstructure as a function of $\omega$~(black lines) and the two frequency ranges of negative mass (red lines). 

As discussed in \S\ref{Sec:negativemass}, sufficient conditions for at least one negative eigen-mass follow from the Gershgorin circle theorem applied to the $2\times 2$ effective mass matrix~\eqref{1mass_effmass}. In particular, from \eqref{Gershgorin} we have 
%------------------------------------------------------
\begin{equation}
    \big|\lambda - M_{jj}\big|\le \big| M_{12}\big|\,,\qquad j=1,2\,,
\end{equation}
%------------------------------------------------------
and an eigen-mass $\lambda$ lies within two Gershgorin discs of the same radius $R=\big| M_{12}\big|$  centered either at $M_{11}$, or $M_{22}$. Thus, sufficient conditions to have one negative eigen-mass is when one of the two Gershgorin discs lies in the negative part of the complex plane ($\mathrm{Re}\lambda<0$), that is $(M_{11}<0\,~\text{and~} |M_{11}|>|M_{12}|)$, or $(
M_{22}<0\,~\text{and~} |M_{22}|>|M_{12}|)$.

%-----------------------------
%-----------------------------
%-----------------------------
%-----------------------------

%-----------------------------
%-----------------------------
\subsubsection{Equivalence with a cell with two micro-masses}\label{Sec:two_micro_masses}

The effective mass matrix of a unit cell with a single micro-mass has a one-to-one correspondence with that with two micro-masses, as will be shown below. Hence, it suffices to consider a single mass-in-mass lattice for studying the effects of microstructure on the effective mass matrix.
Let us consider a single unit cell with macro and micro generalized coordinates $X(t)$, $\Theta(t)$, $x_1(t)$, $\theta_1(t)$, $x_2(t)$, and $\theta_2(t)$ (see Fig.~\ref{Unit-Cell2}).
The balance of micro linear and angular momenta read
%---------------------
\begin{equation}  \label{Micro-Momenta2}
\begin{aligned}
	f_2(t)-f_1(t)+F_{m1}(t) &=m_1\,\ddot{x}_1(t)\,,\\
	t_2(t)-t_1(t)+T_{m1}(t) &=I_{m1}\,\ddot{\theta}_1(t)\,,\\
	f_3(t)-f_2(t)+F_{m2}(t) &=m_2\,\ddot{x}_2(t)\,,\\
	t_3(t)-t_2(t)+T_{m2}(t) &=I_{m2}\,\ddot{\theta}_2(t)\,,
\end{aligned}
\end{equation}
%---------------------  
where $F_{m1}(t)$, $F_{m2}(t)$, and $T_{m1}(t)$, $T_{m2}(t)$ are the applied micro-forces and micro-moments, respectively.
Note that 
%---------------------
\begin{equation} 
	f_j(t)=k\,s_j(t)+k_c\,\psi_j(t)\,,\qquad t_j(t)=k_c\,s_j(t)+k_t\,\psi_j(t)\,,\quad j=1,2,3\,, 
\end{equation}
%---------------------  
where
%---------------------
\begin{equation} \label{constitutive2}
\begin{aligned}
	& s_1(t)=x_1(t)-X(t)\,,&& s_2(t)=x_2(t)-x_1(t)\,,&& s_3(t)=X(t)-x_2(t)\,,   \\
	& \psi_1(t)=\theta_1(t)-\Theta(t)\,,&& \psi_2(t)=\theta_2(t)-\theta_1(t)\,,&& 
	\psi_3(t)=\Theta(t)-\theta_2(t)\,.
\end{aligned}
\end{equation}
%---------------------  
The balance of macro linear and angular momenta read
%---------------------
\begin{equation}  \label{Macro-Momenta2}
\begin{aligned}
	F(t)+f_1(t)-f_3(t) &=M_0\,\ddot{X}(t)\,,\\
	T(t)+t_1(t)-t_3(t) &= I_0\,\ddot{\Theta}(t) \,,
\end{aligned}
\end{equation}
%--------------------- 
where $F(t)$ and $T(t)$ are the applied macro-force and macro-moment, respectively.
Let us assume that the macro and micro forces and torques are harmonic, i.e., $F(t)=\bar{F}\,e^{i\,\omega\,t}$, $T(t)=\bar{T}\,e^{i\,\omega\,t}$, $F_{m1}(t)=\bar{F}_{m1}\,e^{i\,\omega\,t}$, $T_{m1}(t)=\bar{T}_{m1}\,e^{i\,\omega\,t}$, $F_{m2}(t)=\bar{F}_{m2}\,e^{i\,\omega\,t}$, and $T_{m2}(t)=\bar{T}_{m2}\,e^{i\,\omega\,t}$.
This implies that $X(t)=\bar{X}\,e^{i\,\omega\,t}$, $\Theta(t)=\bar{\Theta}\,e^{i\,\omega\,t}$, $x_1(t)=\bar{x}_1\,e^{i\,\omega\,t}$, $\theta_1(t)=\bar{\theta}_1\,e^{i\,\omega\,t}$, $x_2(t)=\bar{x}_2\,e^{i\,\omega\,t}$, and $\theta_2(t)=\bar{\theta}_2\,e^{i\,\omega\,t}$. Thus, the balance equations for the micro and macro variables can be written as
%---------------------
\begin{equation}  \label{Momenta}
\begin{aligned}
	\bar{f}_2-\bar{f}_1(t)+\bar{F}_{m1} &= -\omega^2\,m_1\,\bar{x}_1\,,\\
	\bar{t}_2-\bar{t}_1+\bar{T}_{m1} &= -\omega^2\,I_{m1}\,\bar{\theta}_1\,,\\
	\bar{f}_3-\bar{f}_2+\bar{F}_{m2} &= -\omega^2\,m_2\,\bar{x}_2\,,\\
	\bar{t}_3-\bar{t}_2+\bar{T}_{m2} &= -\omega^2\,I_{m2}\,\bar{\theta}_2\,,\\
	\bar{F}+\bar{f}_1-\bar{f}_3 &= -\omega^2\,M_0\,\bar{X}\,,\\
	\bar{T}+\bar{t}_1-\bar{t}_3 &= -\omega^2\,I_0\,\bar{\Theta} \,.
\end{aligned}
\end{equation}
%---------------------  
Substituting \eqref{constitutive2} into \eqref{Micro-Momenta2} and \eqref{Macro-Momenta2}, one obtains the following matrix equation relating forces and displacements
%---------------------
\begin{equation} \label{F_MA2}
	\begin{bmatrix}
	2k-m_1\,\omega^2 & 2k_c & -k & -k_c & -k & -k_c \\
	2k_c & 2k_t-I_{m1}\,\omega^2  & -k_c & -k_t & -k_c & -k_t \\
	-k & -k_c & 2k-m_2\,\omega^2 & 2k_c & -k & -k_c \\
	-k_c & -k_t & 2k_c & 2k_t-I_{m2}\,\omega^2 & -k_c & -k_t  \\
	-k & -k_c & -k & -k_c & 2k-M_0\,\omega^2 & 2k_c \\
	-k_c & -k_t & -k_c & -k_t & 2k_c & 2k_t-I_0\,\omega^2
	\end{bmatrix}
	\begin{bmatrix}
	\bar{x}_1 \\
	\bar{\theta}_1 \\
	\bar{x}_2 \\
	\bar{\theta}_2 \\
	\bar{X} \\
	\bar{\Theta}
	\end{bmatrix}
	=\begin{bmatrix}
	\bar{F}_{m1} \\
	\bar{T}_{m1} \\
	\bar{F}_{m2} \\
	\bar{T}_{m2} \\
	\bar{F} \\
	\bar{T}
	\end{bmatrix}\,.
\end{equation}
%-----------------------------------------------------------
The above matrix equation can be recast in the inner-outer form as follows. Define the inner and outer generalized displacements 
%------------------------------------------------------
\begin{equation}
    \mathbf{X}_{\text{I}}=
    \begin{bmatrix}
    x_1(t)\\
    \theta_1(t)\\
    x_2(t)\\
    \theta_2(t)
    \end{bmatrix}\,,\qquad 
    \mathbf{X}_{\text{O}}=
    \begin{bmatrix}
    X(t)\\
    \Theta(t)
    \end{bmatrix}\,.
\end{equation}
%------------------------------------------------------
The inner and outer mass matrices read
%------------------------------------------------------
\begin{equation}
    \mathbf{M}_{\text{II}}=
    \begin{bmatrix}
    m & 0 & 0 & 0\\
    0 & I_m & 0 & 0 \\
    0 & 0 & m & 0 \\
    0 & 0 & 0 & I_m
    \end{bmatrix}\,,\qquad
    \mathbf{M}_{\text{OO}}=
    \begin{bmatrix}
   M_0 & 0\\
    0 & I_0
    \end{bmatrix}\,.
\end{equation}
%------------------------------------------------------
The stiffness matrices are written as
%------------------------------------------------------
\begin{equation} 
\mathbf{K}_{\text{OO}}=
    \begin{bmatrix}
   2k & 2k_c\\
    2k_c & 2k_t
    \end{bmatrix}\,,\quad
    \mathbf{K}_{\text{II}}=
    \begin{bmatrix}
    2k & 2k_c  & -k  &  -k_c\\
    2k_c & 2k_t  &  -k_c  &   -k_t \\
    -k & -k_c  & 2k  &  2k_c\\
    -k_c & -k_t  &  2k_c  &   2k_t 
    \end{bmatrix}\,,\quad
    \mathbf{K}_{\text{OI}}=
    -\begin{bmatrix}
    k & k_c  & k & k_c\\
    k_c & k_t & k_c & k_t 
    \end{bmatrix}\,,
\end{equation}
%------------------------------------------------------
and $\mathbf{K}_{\text{OI}}=\mathbf{K}_{\text{IO}}^{\mathsf{T}}$. Equivalently, 
%------------------------------------------------------
\begin{equation}
\mathbf{M}_{\text{OO}}=
    \begin{bmatrix}
   M_0 & 0\\
    0 & I_0
    \end{bmatrix}\,,\qquad\mathbf{M}_{\text{II}}=
    \begin{bmatrix}
    \mathbf{m} & \mathbf{0}\\
    \mathbf{0} & \mathbf{m}
    \end{bmatrix}\,,\qquad  \mathbf{m}=
    \begin{bmatrix}
   m & 0\\
    0 & I_m
    \end{bmatrix}\,,
\end{equation}
%------------------------------------------------------
and
%------------------------------------------------------
\begin{equation} 
\mathbf{K}_{\text{OO}}=2\mathbf{K}\,,\quad
    \mathbf{K}_{\text{II}}=
    \begin{bmatrix}
    2 \mathbf{K} & -\mathbf{K} \\
    -\mathbf{K} & 2 \mathbf{K}
    \end{bmatrix}\,,\quad
    \mathbf{K}_{\text{OI}}=
    -\begin{bmatrix}
    \mathbf{K}   & \mathbf{K}\\ 
    \end{bmatrix}\,,\quad
    \mathbf{K}=
    \begin{bmatrix}
   k & k_c\\
    k_c & k_t
    \end{bmatrix}\,.
\end{equation}
%-------------------------------------------------
The outer and inner force vectors are defined as
%------------------------------------------------------
\begin{equation} 
\mathbf{F}_{\text{O}}=
    \begin{bmatrix}
    F(t) \\
    T(t) 
    \end{bmatrix}\,,\quad\mathbf{F}_{\text{I}}=
    \begin{bmatrix}
    F_{m1} \\
    T_{m1} \\
    F_{m2} \\
    T_{m2}
    \end{bmatrix}\,.
\end{equation}
%------------------------------------------------------
The effective mass is calculated using \eqref{massFourier} and reads
%-------------------------------------------------
\begin{equation}
	\mathbf{M}_{\text{eff}}(\omega)= \begin{bmatrix}
	M_0 +\frac{2 m(k_c^2 -k k_t + I_m k \omega^2)}{ k_c^2 - (k_t - I_m \omega^2) (k - m \omega^2)} 
	& \frac{ 2 I_m k_c m \omega^2 }{k_c^2 - (k_t - I_m \omega^2) (k - m \omega^2)}\\
	\frac{ 2 I_m k_c m \omega^2 }{k_c^2 - (k_t - I_m \omega^2) (k - m \omega^2)} 
	& I_0 +\frac{2 I_m( k_c^2 -k k_t + m k_t \omega^2)}{k_c^2 - (k_t - I_m \omega^2) (k - m \omega^2)}
    \end{bmatrix}\,.
\end{equation}
%-----------------------------------------------------------
Note that the effective mass matrix can be written as a static mass matrix plus an added mass matrix as
\begin{equation}
   \mathbf{M}_{\text{eff}}(\omega)=\mathbf{M}_{\text{static}}+\mathbf{M}_{\text{added}}(\omega)\,,  
\end{equation}
where
%---------------------
\begin{equation} 
	\mathbf{M}_{\text{static}}=\begin{bmatrix}
	M_0 & 0 \\
	0 & I_0
	\end{bmatrix}
	\,,\qquad
	\mathbf{M}_{\text{added}}(\omega)=\begin{bmatrix}
	\widetilde{M}(\omega) & \widetilde{J}(\omega) \\
	\widetilde{J}(\omega) & \widetilde{I}(\omega)
	\end{bmatrix}
	\,,
\end{equation}
%--------------------- 
and 
%---------------------
\begin{equation}\label{Mcoeff2a}  
\begin{aligned}
	\widetilde{M}(\omega) &=\frac{2 m(k_c^2 -k k_t + I_m k \omega^2)}{ k_c^2 
	- (k_t - I_m \omega^2) (k - m \omega^2)}   \,,\\
	\widetilde{J}(\omega) &=  \frac{ 2 I_m k_c m \omega^2 }{k_c^2 - (k_t - I_m \omega^2) (k - m \omega^2)}  \,,\\
	\widetilde{I}(\omega) &= \frac{2 I_m( k_c^2 -k k_t + m k_t \omega^2)}{k_c^2 
	- (k_t - I_m \omega^2) (k - m \omega^2)}\,.
\end{aligned}
\end{equation}
%-----------------------------------------------------------
Comparing with the effective mass matrix of the single micro-mass cell in \eqref{1mass_effmass} we note the following equivalence
%-----------------------------------------------------------
\begin{equation}
	\mathbf{M}^{2-\text{mass}}_{\text{eff}}(2k,2k_c,2k_t,m,I_m,2M_0,2I_0)=2\mathbf{M}^{1-\text{mass}}_{\text{eff}}(k,k_c,k_t,m,I_m,M_0,I_0)\,.
\end{equation}
%-----------------------------------------------------------
Thus, a single micro-mass cell with micro-mass $m$ and micro-inertia $I_m$ is equivalent to a double micro-mass cell with the same micro-mass and micro-inertia, doubled outer stiffness and macro-mass and inertia. There is a similar equivalence relation for a unit cell with $N$ micro-masses in series. Thus, hereafter it suffices to consider a microstructure  with a single mass-in-mass unit cell~(see top-left panel of Fig.~\ref{Unit-Cell2}).

%-----------------------------------------------------------
%-----------------------------------------------------------
\subsection{Bloch analysis and frequency bands}\label{Sec:Blochanalysis}

\begin{figure}[t!]
\centering
\includegraphics[width=0.95\textwidth]{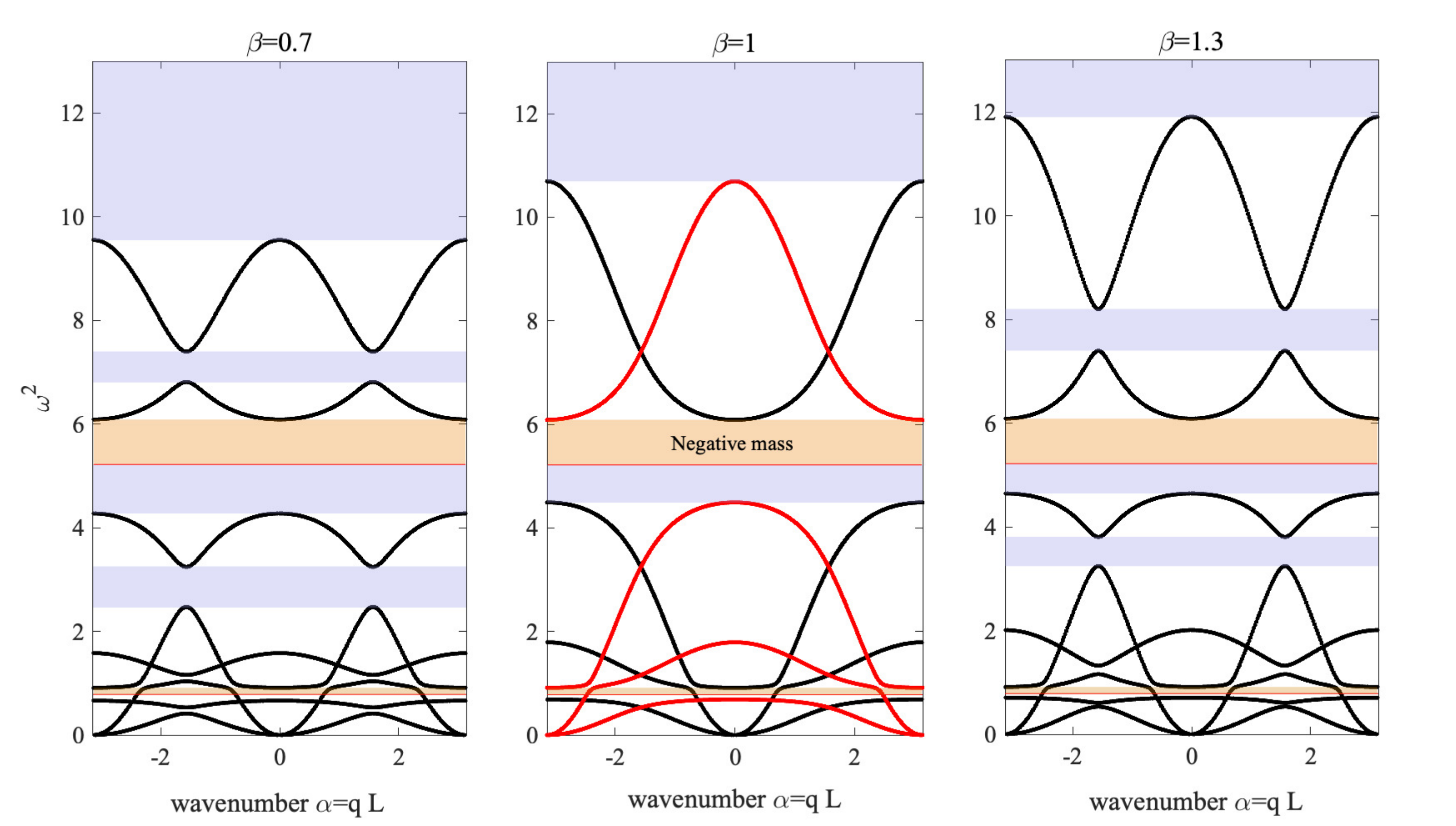}
\vspace*{0.10in}
\centering
\vskip 0.4in
\includegraphics[width=0.85\textwidth]{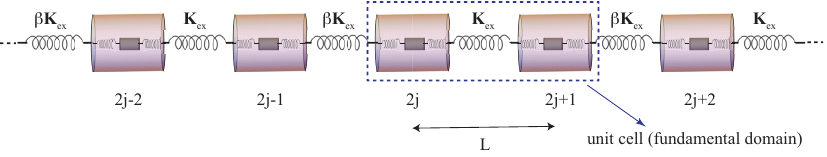}
\vskip 0.3in
\includegraphics[width=0.85\textwidth]{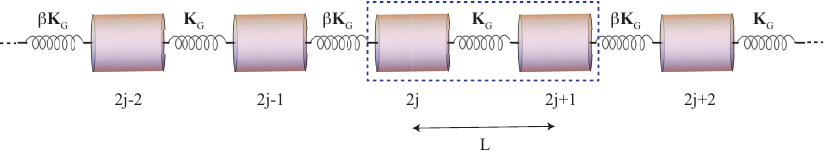}
\vspace*{0.10in}
\caption{Bottom panels: (top) a periodic $1$D composite lattice with alternating nearest-neighbor cell-cell stiffness matrices and (bottom) the associated reduced~(condensed) lattice. Top panel: Frequency dispersion bands of the $1$D composite lattice for $\beta=0.7,1, 1.3$. The (black) symmetric and (red) anti-symmetric bands are also depicted for the simple lattice $\beta=1$. Bandgaps (light violet) and frequency range of negative mass (ochre) are also shown. Natural frequencies of the microstructure (red lines): $\omega_1^2=5.22,\omega_2^2=0.78$.}
\label{alternatinge_latticebandgaps}
\end{figure}

In this section, we study the spectral properties of the $1$D composite lattice depicted in the second panel from the bottom of Fig.~\ref{alternatinge_latticebandgaps}. The unit cell (fundamental domain) of the composite lattice has two mass-in-mass cells that are separated by distance $L$ and are connected to each other by an outer chiral spring with elastic constants $K$, $K_c$, and $K_t$. Each mass-in-mass cell is identical to that shown in the top-left panel of Fig.~\ref{Unit-Cell2}.
In order to avoid unstable modes, it is assumed that $kk_t-k_c^2>0$, and $KK_t-K_c^2>0$. Note that $k,k_t,K,K_t>0$.
Let us define the following $4\times 4$~local stiffness matrices
%------------------------------------------------------
\begin{equation}\label{KGs}
   \mathbf{K}_{\text{ex}}=
    \begin{bmatrix}
    \mathbf{K}_{\text{G}} & \mathbf{0}\\
    \mathbf{0} & \mathbf{0}
    \end{bmatrix}\,,\qquad 
    \mathbf{K}_{\text{G}}=
    \begin{bmatrix}
   K & K_c\\
    K_c & K_t
    \end{bmatrix}\,,
\end{equation}
%------------------------------------------------------
where $\mathbf{K}_{\text{G}}$ and the null matrix $\mathbf{0}$ are $2\times 2$ matrices. 
The generic cells $2j$ and $2j+1$ interact via the stiffness matrix $\mathbf{K}_{\text{ex}}$ while the nearest-neighbor unit cells interact through the stiffness matrix $\beta\mathbf{K}_{\text{ex}}$, with $\beta>0$. The case of a simple lattice is recovered when $\beta=1$. 

%-----------------------------
%-----------------------------
%-----------------------------
%-----------------------------

We next perform a dynamic Bloch analysis of the infinite lattice, whose unit cell has $4$ degrees of freedom~($2$~outer and $2$~inner generalized displacements). The frequency bands follow by solving a linear eigenvalue problem in the squared frequency $\omega^2$~(c.f., e.g., \cite{FedeleDefects2005}, and references therein). Then, we consider the associated reduced (effective) lattice by lumping the microstructure of each cell to a single mass with the $2\times 2$ effective mass matrix  $\mathbf{M}_{\text{eff}}(\omega)$ in \eqref{1mass_effmass}. The reduced~(condensed) lattice has unit cells with $2$ degrees of freedom~($2$~outer generalized displacements only). The frequency bands follow by solving a smaller~$2\times 2$, but nonlinear eigenvalue problem in $\omega^2$ because the effective mass matrix is frequency dependent. It is shown that the full and reduced lattice models give the same frequency bands. On the one hand, the full lattice has a stiffness matrix double the size of that of the reduced lattice, but the Bloch eigenvalue problem of the former is linear. On the other hand, the reduced (effective) lattice yields a nonlinear Bloch eigenvalue problem of a smaller size. 

%------------------------------------------------------
%------------------------------------------------------
\subsubsection{The full lattice model}

From \S\ref{Sec:Lattice}, the Fourier transform of the associated dynamical equations~\eqref{Micro-Momenta}~and~\eqref{Macro-Momenta} can be written in the compact matrix form of \eqref{IOmatrixsys} as $-\omega^2 \mathbf{M}_{\text{cell}}\hat{\mathbf{u}} + \mathbf{K}_{\text{cell}}\hat{\mathbf{u}}=\hat{\mathbf{F}}$, where 
%------------------------------------------------------
\begin{equation}\label{IOmatrixsys3}
   \mathbf{M}_{\text{cell}}=\begin{bmatrix}
\mathbf{M}_{\text{OO}} & \mathbf{0}\\
    \mathbf{0} & \mathbf{M}_{\text{II}}
    \end{bmatrix}\,,\quad
    \mathbf{K}_{\text{cell}}=
    \begin{bmatrix}
    \mathbf{K}_{\text{OO}} & \mathbf{K}_{\text{OI}}\\
    \mathbf{K}_{\text{IO}} & \mathbf{K}_{\text{II}}
    \end{bmatrix}=\begin{bmatrix}
    2\mathbf{K} & -2\mathbf{K}\\
    -2\mathbf{K} & 2\mathbf{K}
    \end{bmatrix}\,,\quad
    \hat{\mathbf{u}}=
    \begin{bmatrix}
    \hat{\mathbf{X}}_{\text{O}}\\
   \hat{\mathbf{X}}_{\text{I}}
    \end{bmatrix}\,,\quad 
    \hat{\mathbf{F}}=
    \begin{bmatrix}
    \hat{\mathbf{F}}_{\text{O}}\\
    \hat{\mathbf{F}}_{\text{I}}
    \end{bmatrix}\,. 
\end{equation}
%------------------------------------------------------
Here, the inner and outer mass matrices are given in~\eqref{massmatrices} and the stiffness matrix $\mathbf{K}$ is defined in~\eqref{stiffmatrixK}. 
The vector $\hat{\mathbf{u}}\in \mathbb{C}^4$ lists the $2\times 1$ vectors of the Fourier amplitudes of the outer and inner generalized displacements $(\hat{\mathbf{X}}_{\text{O}},\hat{\mathbf{X}}_{\text{I}})$ given in \eqref{Fouriergendisp}. The outer and inner forces  $(\hat{\mathbf{F}}_{\text{O}},\hat{\mathbf{F}}_{\text{I}})$ are defined in \eqref{forcingC}. 
In the following we will assume that the inner forces vanish, i.e., $\hat{\mathbf{F}}_{\text{I}}=\mathbf{0}$. 

The location of each cell of the fundamental domain shown in Fig.~\ref{alternatinge_latticebandgaps} is $X_j=j L$ for $j=1,\hdots N$. The even cells $2j$ are connected to their nearest-neighbor unit cells $2j+1$ and $2j-1$ by springs with stiffness matrices $\mathbf{K}_{\text{ex}}$ and $\beta \mathbf{K}_{\text{ex}}$ ($\beta>0$), respectively.
Let us assume harmonic loads. The displacements of odd and even unit cells are assumed to be of the following forms
%------------------------------------------------------
\begin{equation}\label{disps}
    \hat{\mathbf{u}}_{2j}=\hat{\mathbf{u}}_0 \mathrm{e}^{i q X_{2j}}=\hat{\mathbf{u}}_0 \mathrm{e}^{i \alpha 2 j}\,,\qquad \hat{\mathbf{u}}_{2j+1}=\hat{\mathbf{u}}_1 \mathrm{e}^{i q X_{2j+1}}=\hat{\mathbf{u}}_1 \mathrm{e}^{i \alpha (2 j+1)}\,,
\end{equation}
%------------------------------------------------------
where $\hat{\mathbf{u}}_0,\hat{\mathbf{u}}_1\in \mathbb{C}^4$, and $\alpha=q L$. Note that such a form for the solutions follows from the block-circulant nature of the stiffness and mass matrices of the lattice, which allows them to be block-diagonalized by the block form of the discrete Fourier transform matrix (cf., e.g., \cite{pratapa2018bloch}). The dynamical governing equations then read
%------------------------------------------------------
\begin{equation}\label{balances}
\begin{aligned}
       -\beta \mathbf{K}_{\text{ex}} \hat{\mathbf{u}}_{2j-1}+[(1+\beta)\mathbf{K}_{\text{ex}}+\mathbf{K}_{\text{cell}}]\hat{\mathbf{u}}_{2j}-\mathbf{K}_{\text{ex}} \hat{\mathbf{u}}_{2j+1}-\omega^2 \mathbf{M}_{\text{cell}}\hat{\mathbf{u}}_{2j}&=\mathbf{0}\,,
   \\
    -\mathbf{K}_{\text{ex}} \hat{\mathbf{u}}_{2j}+[(1+\beta)\mathbf{K}_{\text{ex}}+\mathbf{K}_{\text{cell}}] \hat{\mathbf{u}}_{2j+1}-\beta \mathbf{K}_{\text{ex}} \hat{\mathbf{u}}_{2j+2}-\omega^2 \mathbf{M}_{\text{cell}}\hat{\mathbf{u}}_{2j+1}&=\mathbf{0}\,.
\end{aligned}
\end{equation}
%------------------------------------------------------
Using \eqref{disps}, one obtains
%------------------------------------------------------
\begin{equation}\label{balance2s}
\begin{aligned}
       [(1+\beta)\mathbf{K}_{\text{ex}}+\mathbf{K}_{\text{cell}}] \hat{\mathbf{u}}_{0}\mathrm{e}^{i \alpha 2 j}-\mathbf{K}_{\text{ex}} \hat{\mathbf{u}}_{1}\mathrm{e}^{i \alpha (2 j+1)}-\beta \mathbf{K}_{\text{ex}} \hat{\mathbf{u}}_{1}\mathrm{e}^{i \alpha (2 j-1)}-\omega^2 \mathbf{M}_{\text{cell}}\hat{\mathbf{u}}_{0}\mathrm{e}^{i \alpha 2 j} &=\mathbf{0}\,,
   \\
    [(1+\beta)\mathbf{K}_{\text{ex}}+\mathbf{K}_{\text{cell}}]  \hat{\mathbf{u}}_{1}\mathrm{e}^{i \alpha (2 j+1)}-\beta \mathbf{K}_{\text{ex}} \hat{\mathbf{u}}_{0}\mathrm{e}^{i \alpha (2 j+2)}-\mathbf{K}_{\text{ex}} \hat{\mathbf{u}}_{0}\mathrm{e}^{i \alpha 2j}-\omega^2 \mathbf{M}_{\text{cell}}\hat{\mathbf{u}}_{1}\mathrm{e}^{i \alpha (2 j+1)} &=\mathbf{0}\,.
\end{aligned}
\end{equation}
%------------------------------------------------------
Dividing the first and second equations by $\mathrm{e}^{i \alpha 2 j}$ and $\mathrm{e}^{i \alpha (2 j+1)}$, respectively, one obtains
%------------------------------------------------------
\begin{equation}\label{balance3s}
\begin{aligned}
       [(1+\beta)\mathbf{K}_{\text{ex}}+\mathbf{K}_{\text{cell}}] \hat{\mathbf{u}}_{0}-a\mathbf{K}_{\text{ex}} \hat{\mathbf{u}}_{1}-\omega^2 \mathbf{M}_{\text{cell}}\hat{\mathbf{u}}_{0} &=\mathbf{0}\,,
   \\
    [(1+\beta)\mathbf{K}_{\text{ex}}+\mathbf{K}_{\text{cell}}]\hat{\mathbf{u}}_{1}-a^{\dagger}\mathbf{K}_{\text{ex}} \hat{\mathbf{u}}_{0}-\omega^2 \mathbf{M}_{\text{cell}}\hat{\mathbf{u}}_{1}&=\mathbf{0}\,,
\end{aligned}
\end{equation}
%------------------------------------------------------
where $a=\mathrm{e}^{i \alpha }+ \beta \mathrm{e}^{-i\alpha }$, and $a^{\dagger}$ is the complex conjugate of $a$.  This can be rewritten in the following compact form
%------------------------------------------------------
\begin{equation}\label{Balance5s}
\begin{bmatrix}
(1+\beta)\mathbf{K}_{\text{ex}}+\mathbf{K}_{\text{cell}} - \omega^2\mathbf{M}_{\text{cell}} & -a\mathbf{K}_{\text{ex}}\\ -a^{\dagger}\mathbf{K}_{\text{ex}} & (1+\beta)\mathbf{K}_{\text{ex}}+\mathbf{K}_{\text{cell}} - \omega^2\mathbf{M}_{\text{cell}}
\end{bmatrix}
\begin{bmatrix}
\hat{\mathbf{u}}_0 \\ 
\hat{\mathbf{u}}_1
\end{bmatrix} =
\begin{bmatrix}
\mathbf{0} \\ 
\mathbf{0}
\end{bmatrix}\,.
\end{equation}
%------------------------------------------------------
Note that the coefficient matrix is Hermitian, and hence, has real eigenvalues. The dispersion bands $\omega^2(\alpha)$ can be evaluated by solving the $8\times 8$ linear eigenvalue problem above by imposing the vanishing of the determinant of the coefficient matrix. This yields an $8$\emph{th}-order polynomial equation in $\omega^2$, whose roots yield $8$ frequency bands.

%------------------------------------------------------
%------------------------------------------------------
\subsubsection{The reduced lattice model via the effective mass} 

We next lump the microstructure of each cell to a single mass with the $2\times 2$ effective mass matrix  $\mathbf{M}_{\text{eff}}(\omega)$ in \eqref{1mass_effmass} related to the outer Fourier amplitude displacements $\hat{\mathbf{v}}=\hat{\mathbf{X}}_{\text{O}}\in \mathbb{C}^2$  given in \eqref{Fouriergendisp}$_1$. The reduced~(condensed) lattice has unit cells with $2$ degrees of freedom~(outer displacements of translation and twist) and it is depicted in the bottom panel of Fig.~\ref{alternatinge_latticebandgaps}. In the following we will show that the frequency bands are obtained by solving a $4\times 4$ nonlinear eigenvalue problem in $\omega^2$ as the effective mass matrix is frequency dependent. 

The effective mass cells are connected via outer chiral elastic springs characterized by the $2\times2$ stiffness matrix $\mathbf{K}_G$~(see \eqref{KGs}$_2$). In particular, even nodes interact with their  successive (odd) node by a spring of stiffness $\mathbf{K}_G$ and interact with their preceding (odd) node by a spring of stiffness $\beta \mathbf{K}_G$. Odd nodes are connected to their successive (even node) by a spring of stiffness $\beta \mathbf{K}_G$ and to their preceding (even) node by a spring with stiffness $\mathbf{K}_G$. For a harmonic motion the associated Fourier amplitude of the outer displacements~(translation and twist) of odd and even nodes are assumed to be of the following form
%------------------------------------------------------
\begin{equation}\label{disp}
    \hat{\mathbf{v}}_{2j}=\hat{\mathbf{v}}_0 \mathrm{e}^{i q X_{2j}}=\hat{\mathbf{v}}_0 \mathrm{e}^{i \alpha 2 j}\,,\qquad \hat{\mathbf{v}}_{2j+1}=\hat{\mathbf{v}}_1 \mathrm{e}^{i q X_{2j+1}}=\hat{\mathbf{v}}_1 \mathrm{e}^{i \alpha (2 j+1)}\,,
\end{equation}
%------------------------------------------------------
where $\hat{\mathbf{v}}_0,\hat{\mathbf{v}}_1\in \mathbb{C}^2$, and $\alpha=q L$. The dynamical equations of even and odd node masses are written as
%------------------------------------------------------
\begin{equation}\label{balance}
\begin{aligned}
       (1+\beta)\mathbf{K}_G \hat{\mathbf{v}}_{2j}-\mathbf{K}_G \hat{\mathbf{v}}_{2j+1}-\beta \mathbf{K}_G \hat{\mathbf{v}}_{2j-1}-\omega^2 \mathbf{M}_{\text{eff}}(\omega)\hat{\mathbf{v}}_{2j} &=\mathbf{0}\,,
   \\
     (1+\beta)\mathbf{K}_G \hat{\mathbf{v}}_{2j+1}-\beta \mathbf{K}_G \hat{\mathbf{v}}_{2j+2}-\mathbf{K}_G \hat{\mathbf{v}}_{2j}-\omega^2 \mathbf{M}_{\text{eff}}(\omega)\hat{\mathbf{v}}_{2j+1}&=\mathbf{0}\,.
\end{aligned}
\end{equation}
%------------------------------------------------------
Using \eqref{disp}, one writes
%------------------------------------------------------
\begin{equation}\label{balance2}
\begin{aligned}
        (1+\beta)\mathbf{K}_G \hat{\mathbf{v}}_{0}\mathrm{e}^{i \alpha 2 j}-\mathbf{K}_G \hat{\mathbf{v}}_{1}\mathrm{e}^{i \alpha (2 j+1)}-\beta \mathbf{K}_G \hat{\mathbf{v}}_{1}\mathrm{e}^{i \alpha (2 j-1)}-\omega^2 \mathbf{M}_{\text{eff}}(\omega)\hat{\mathbf{v}}_{0}\mathrm{e}^{i \alpha 2 j}&=\mathbf{0}\,,
   \\
    (1+\beta)\mathbf{K}_G \hat{\mathbf{v}}_{1}\mathrm{e}^{i \alpha (2 j+1)}-\beta \mathbf{K}_G \hat{\mathbf{v}}_{0}\mathrm{e}^{i \alpha (2 j+2)}-\mathbf{K}_G \hat{\mathbf{v}}_{0}\mathrm{e}^{i \alpha 2j}-\omega^2 \mathbf{M}_{\text{eff}}(\omega)\hat{\mathbf{v}}_{1}\mathrm{e}^{i \alpha (2 j+1)}&=\mathbf{0}\,.
\end{aligned}
\end{equation}
%------------------------------------------------------
Dividing the first and second equations by $\mathrm{e}^{i \alpha 2 j}$ and $\mathrm{e}^{i \alpha (2 j+1)}$, respectively, one obtains
%------------------------------------------------------
\begin{equation}\label{balance3}
\begin{aligned}
       (1+\beta)\mathbf{K}_G \hat{\mathbf{v}}_{0}-a\mathbf{K}_G \hat{\mathbf{v}}_{1}-\omega^2 \mathbf{M}_{\text{eff}}(\omega)\hat{\mathbf{v}}_{0}&=\mathbf{0}\,,
   \\
    (1+\beta)\mathbf{K}_G \hat{\mathbf{v}}_{1}-a^{\dagger}\mathbf{K}_G \hat{\mathbf{v}}_{0}-\omega^2 \mathbf{M}_{\text{eff}}(\omega)\hat{\mathbf{v}}_{1}&=\mathbf{0}\,,
\end{aligned}
\end{equation}
%------------------------------------------------------
where $a=\mathrm{e}^{i \alpha }+ \beta \mathrm{e}^{-i\alpha }$, and $a^{\dagger}$ is the complex conjugate of $a$.  This can be written in the following compact form
%------------------------------------------------------
\begin{equation}\label{Balance4}
\left( \begin{bmatrix}
(1+\beta)\mathbf{K}_{\mathrm{G}} & -a\mathbf{K}_{\mathrm{G}}\\ -a^{\dagger}\mathbf{K}_{\mathrm{G}} & (1+\beta)\mathbf{K}_{\mathrm{G}}
\end{bmatrix}-\omega^2 
\begin{bmatrix}
\mathbf{M}_{\text{eff}}(\omega) & \mathbf{0}\\ \mathbf{0} & \mathbf{M}_{\text{eff}}(\omega)
\end{bmatrix}\right) 
\begin{bmatrix}
\hat{\mathbf{v}}_0 \\ 
\hat{\mathbf{v}}_1
\end{bmatrix} =
\begin{bmatrix}
\mathbf{0} \\ 
\mathbf{0}
\end{bmatrix}\,.
\end{equation}
%------------------------------------------------------
Thus, we obtain the following $4\times 4$ nonlinear eigenvalue problem in $\omega^2$
%------------------------------------------------------
\begin{equation}\label{Balance5}
\begin{bmatrix}
(1+\beta)\mathbf{K}_{\mathrm{G}} - \omega^2\mathbf{M}_{\text{eff}}(\omega) & -a\mathbf{K}_{\mathrm{G}}\\ -a^{\dagger}\mathbf{K}_{\mathrm{G}} & (1+\beta)\mathbf{K}_{\mathrm{G}} - \omega^2\mathbf{M}_{\text{eff}}(\omega)
\end{bmatrix}
\begin{bmatrix}
\hat{\mathbf{v}}_0 \\ 
\hat{\mathbf{v}}_1
\end{bmatrix} =
\begin{bmatrix}
\mathbf{0} \\ 
\mathbf{0}
\end{bmatrix}\,, 
\end{equation}
%------------------------------------------------------
because the effective mass matrix is frequency dependent~(see \eqref{1mass_effmass}). The dispersion bands $\omega^2(\alpha)$ follow by imposing the vanishing of the determinant of the above $4\times 4$ coefficient matrix. This gives the same $8$\emph{th}-order polynomial equation in $\omega^2$ that follows from the full linear eigenvalue problem in \eqref{Balance5s}.  
The nonlinear eigenvalue problem~\eqref{Balance5} reduces to two smaller $2\times 2$~nonlinear eigenvalue problems for the simple lattice with $\beta=1$ as is amenable to algebraic simplification.

\begin{remark}
In the case of an uniform lattice, i.e., $\beta=1$, \eqref{Balance5} reduces to 
%------------------------------------------------------
\begin{equation}\label{Balance15}
\begin{bmatrix}
2\mathbf{K}_{\mathrm{G}} - \omega^2\mathbf{M}_{\text{eff}}(\omega) & -2\cos\alpha\,\mathbf{K}_{\mathrm{G}}\\ -2\cos\alpha\,\mathbf{K}_{\mathrm{G}} & 2\mathbf{K}_{\mathrm{G}} - \omega^2\mathbf{M}_{\text{eff}}(\omega)
\end{bmatrix}
\begin{bmatrix}
\hat{\mathbf{v}}_0 \\ 
\hat{\mathbf{v}}_1
\end{bmatrix} =
\begin{bmatrix}
\mathbf{0} \\ 
\mathbf{0}
\end{bmatrix}\,.
\end{equation}
%------------------------------------------------------
The associated matrix is real-symmetric. The mirror-symmetry of the uniform lattice, or equivalently, rotational symmetry with an angle of $\pi$, permits solutions of the system~\eqref{Balance15} as either ($+$)~symmetric modes $[\hat{\mathbf{v}}_0,\,\hat{\mathbf{v}}_0]$~($\hat{\mathbf{v}}_1=\hat{\mathbf{v}}_0$) or ($-$)~anti-symmetric modes $[\hat{\mathbf{v}}_0,\, -\hat{\mathbf{v}}_0]$~$(\hat{\mathbf{v}}_1=-\hat{\mathbf{v}}_0)$ (cf., e.g., \cite{sharma2021real}, and references therein). For symmetric modes the $4\times 4$ matrix equation in~\eqref{Balance15} reduces to two identical $2\times 2$ matrix equations for $\hat{\mathbf{v}}_0$: $\left[2(1-\cos\alpha) \mathbf{K}_{\mathrm{G}} - \omega^2\mathbf{M}_{\text{eff}}(\omega)\right]\hat{\mathbf{v}}_0=\mathbf{0}$.
For the anti-symmetric modes we instead have $\left[2(1+\cos\alpha) \mathbf{K}_{\mathrm{G}} - \omega^2\mathbf{M}_{\text{eff}}(\omega)\right]\hat{\mathbf{v}}_0=\mathbf{0}$.
In compact form, the symmetric and anti-symmetric modes satisfy the following $2\times 2$ nonlinear eigenvalue problem
%------------------------------------------------------
\begin{equation}\label{eqbeta1}
    (b^{\pm} \mathbf{K}_{\mathrm{G}} - \omega^2\mathbf{M}_{\text{eff}}(\omega))\hat{\mathbf{v}}_0=\mathbf{0}\,, 
\end{equation}
%------------------------------------------------------
where $b^{\pm}=2(1\mp \cos\alpha)$.  Non-trivial solutions for symmetric ($+$) and anti-symmetric ($-$) modes exist if the determinant of the linear system in~\eqref{eqbeta1} vanishes. Such a condition provides the squared eigenfrequencies $\omega^2$ by solving the following nonlinear equation 
%------------------------------------------------------
\begin{equation}
    \left|b^{\pm}\mathbf{K_{\text{G}}}-\omega^{2}\mathbf{M_{\text{eff}}(\omega)}\right|=0\,,
\end{equation}
%------------------------------------------------------
or 
%------------------------------------------------------
\begin{equation}
\begin{vmatrix}
    b^{\pm} K-\omega^{2}M_{11}(\omega) & b^{\pm} K_{c}-\omega^{2}M_{12}(\omega)\\
    b^{\pm} K_{c}-\omega^{2}M_{21}(\omega) &  b^{\pm} K_{t}-\omega^{2}M_{22}(\omega)
\end{vmatrix}=0\,.
\end{equation}
%------------------------------------------------------
This simplifies to $Q_1(\omega)\,\omega^4 + Q_2(\omega)\,\omega^2+Q_3=0$, with 
%------------------------------------------------------
\begin{equation}
\begin{aligned}
    Q_1 &= \big|\mathbf{M_{\text{eff}}}\big|=-M_{12}^2(\omega) + M_{11}(\omega) M_{22}(\omega)\,, 
    \\
     Q_2 &= b^{\pm} (-K_t M_{11}(\omega) + 2 K_c M_{12}(\omega) - K M_{22}(\omega))\,,
    \\
    Q_3 &= (b^{\pm})^2 \big|\mathbf{K_{\text{G}}}\big|=- (b^{\pm})^2 (K_c^2 - K K_t)\,,
\end{aligned}
\end{equation}
%------------------------------------------------------
where use has been made of the symmetry $M_{12}=M_{21}$. 
This reduces to the following $4$\emph{th}-order polynomial in $r=\omega^2$:
%------------------------------------------------------
\begin{equation}\label{4thorder}
     A r^4 + B r^3 + C r^2 + D r +E=0\,,
\end{equation}
%------------------------------------------------------
where
%------------------------------------------------------
\begin{equation}
\begin{aligned}
    A=& -I_0 I_m m M_0\,,
    \\
    B=& b^{\pm} I_m m (I_0 K + K_t M_0) + 
 2 (I_0 k_t m M_0 + I_m k_t m M_0 + I_0 I_m k (m + M_0))\,,
    \\
    C=& (b^{\pm})^2 I_m (K_c^2 - K K_t) m + 4 (I_0 + I_m) (k_c^2 - k k_t) (m + M_0) - 
    \\
 & 2 b^{\pm} (I_0 K (I_m k + k_t m) + K_t k_t m M_0 + I_m (-2 k_c K_c m + K k_t m + k K_t (m + M_0)))\,,
      \\
     D=& 2 b^{\pm} (-b^{\pm} (K_c^2 - K K_t)( I_m k + k_t m)  - 2 (k_c^2 - k k_t) ((I_0 + I_m) K + K_t (m + M_0))) \,,
   \\
   E= & -4 (b^{\pm})^2 (k_c^2 - k k_t) (K_c^2 - K K_t)\,.
\end{aligned}
\end{equation}
%------------------------------------------------------
Thus, we have $4$ bands of symmetric modes~($b^{+}$) and $4$ bands of anti-symmetric modes~($b^{-}$). 
The associated eigenvector $\mathbf{u}_0$ follows from~\eqref{eqbeta1}. For example, 
%------------------------------------------------------
\begin{equation}
(\hat{\mathbf{v}}_0)_1=R,\qquad (\hat{\mathbf{v}}_0)_2=1\,,\qquad R=-\frac{b^{\pm} K_{c}-\omega^{2}M_{12}(\omega)}{b^{\pm} K-\omega^{2}M_{11}(\omega)}\,,
\end{equation}
%------------------------------------------------------
if the denominator of $R$ does not vanish.
%------------------------------------------------------

Zero-frequency, or energy modes exist if $E=0$, that is $-4 (b^{\pm})^2 (k_c^2 - k k_t) (K_c^2 - K K_t)=0$.
This vanishes if either i) $b^{\pm}=0$, i.e., $\cos\alpha=\pm 1$, or ii) $k_c^2 - k k_t=0$, or $K_c^2 - K K_t=0$,
which prescribes the value of the chiral coefficients so that  either the determinant of the inner stiffness matrix $\mathbf{K}$ or that of the outer stiffness matrix $\mathbf{K_G}$ vanishes. For example, from i) zero-frequency modes satisfy  $\alpha=q L=0$, and $q L=-\pi,0,\pi$. Such modes correspond to generalized rigid displacements~(translation and twist) of the entire chain ($q L=0$) or to opposite displacements at even and odd nodes ($q L=\pm \pi$). 
Such modes are `floppy' in the sense that they are \emph{mechanisms} that do not require any deformation energy, or force.
An example is
%-----------------------------------------------------
\begin{equation}
    \hat{\mathbf{v}}_{\text{floppy}}=\begin{bmatrix}
    \bar{X}  \\
    \bar{\Theta} 
    \end{bmatrix}=\begin{bmatrix}
    -\sqrt{\frac{K_t}{K}}\\
    1
    \end{bmatrix}\,.
\end{equation}
%-----------------------------------------------------
\end{remark}

\paragraph{Examples} 
Consider the $1$D composite lattice and its reduced model depicted in the bottom panels of Fig.~\ref{alternatinge_latticebandgaps}. The inner parameters $m=0.25, I_m=0.25, k=1/2, k_t=1/4$, and the outer parameters $M_0=1.5, I_0=1.5, K=2, K_t=2$ are chosen. The inner and outer chiral stiffnesses are selected to avoid zero-energy modes, in particular $k_c=\gamma\sqrt{k k_t}$, and $K_c=\Gamma\sqrt{K K_t}$, where $|\gamma|,|\Gamma|<1$. When $|\gamma|,|\Gamma|=1$ zero-energy modes exist. We choose $\gamma=\Gamma=0.7$.  We consider the reduced~(condensed) lattice and frequency dispersion bands $\omega^2(\alpha)$ are computed by solving the nonlinear eigenvalue problem in $\omega^2$ in \eqref{Balance5}, which is an $8${\emph{th}}-order polynomial in $\omega^2$. The bisection method is used to find the $8$ roots of the $4\times 4$~determinant of matrix system~\eqref{Balance5}, which are identical to those obtained by solving the $8\times 8$~linear eigenvalue problem~\eqref{Balance5s} of the full lattice.  

The frequency bands for $\beta=0.7,1,1.3$ are depicted in the top panels of Fig.~\ref{alternatinge_latticebandgaps}. The bandgap region of negative mass is also shown. For the uniform lattice~($\beta=1$) the roots of the quartic algebraic equation~\eqref{4thorder} yields the frequency bands of symmetric modes~(black lines) and anti-symmetric modes~(red lines), the only two modes permitted by the mirror-symmetry of the lattice. Two pairs of frequency bands intersect at special points indicating the existence of $1$D analogues of Dirac cones~\citep{Dirac_cones_Ochiai_Onada2009,Dirac_coneHe_Chan2005}. The emergence of such special points is attributed to band crossing protected by the mirror-symmetry of the uniform lattice~\citep{Dirac_coneHe_Chan2005}. The Dirac nodes unbuckle for the cases $\beta\neq 1$ because the mirror-symmetry of the lattice is broken by the alternating outer springs of different stiffness along the lattice. A bandgap opens for each band pair, where wave propagation is forbidden. The composite lattice becomes the $1$D analogue of a Chern, or topological insulator~\citep{Vitelli_topologicalmatter}.

%-----------------------------
%-----------------------------
\begin{figure}[t!]
\centering
\begin{subfigure}[b]{0.8\textwidth}
   \includegraphics[width=1\linewidth]{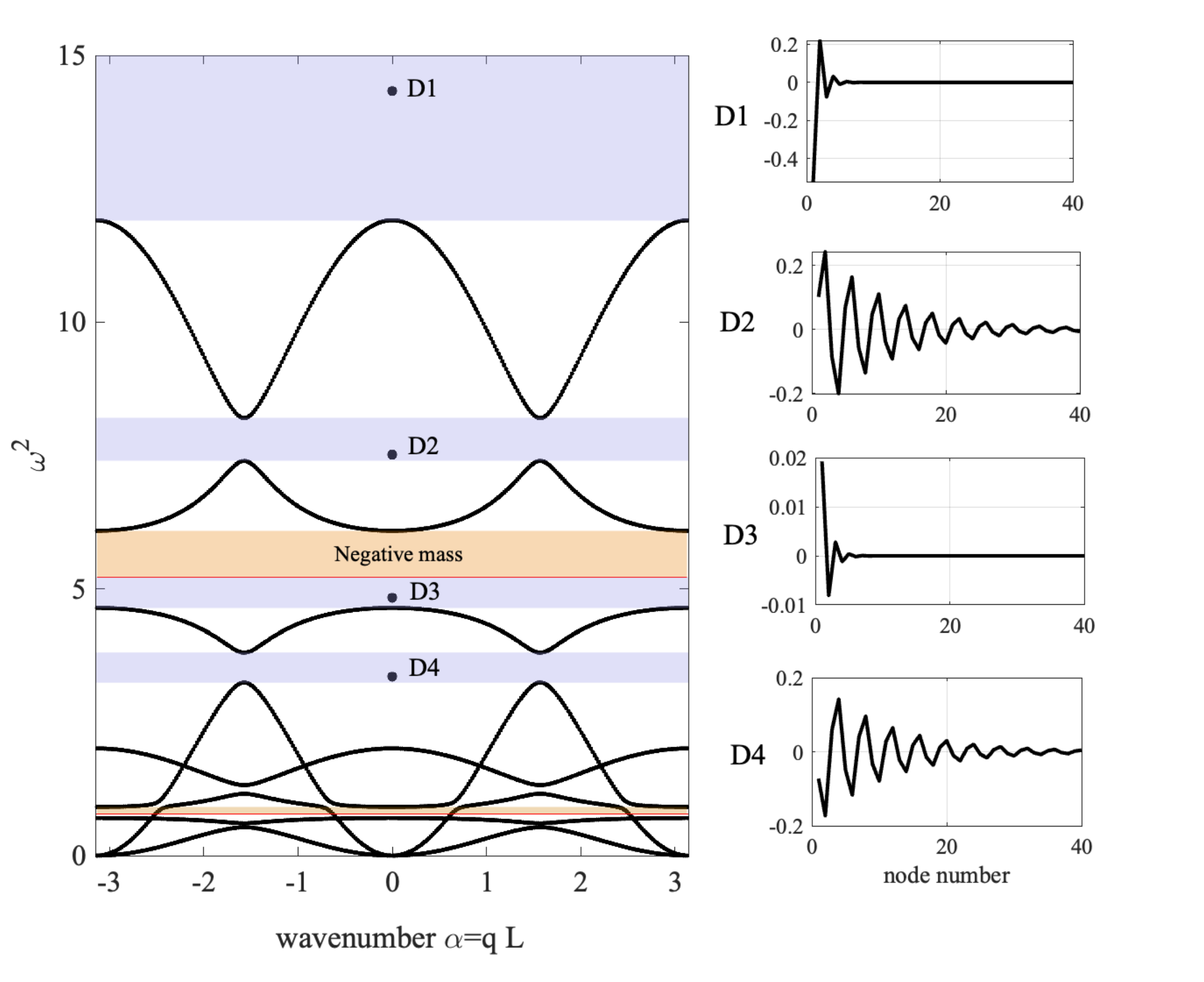}
\end{subfigure}
%-----------------------------
\vskip 0.3in
\begin{subfigure}[b]{0.9\textwidth}
   \includegraphics[width=1\linewidth]{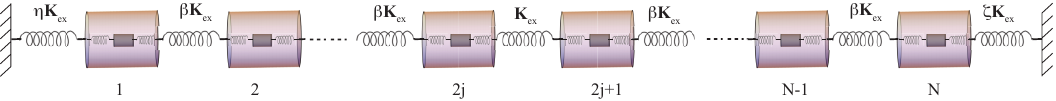}
\end{subfigure}
%-----------------------------
\vskip 0.2in
\begin{subfigure}[b]{0.9\textwidth}
   \includegraphics[width=1\linewidth]{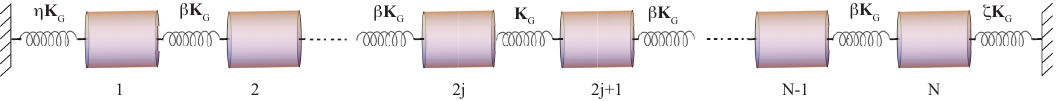}
\end{subfigure}
%-----------------------------
\caption{Bottom panels: (top) a $1$D finite composite lattice with alternating nearest-neighbor cell-cell stiffness matrices~($N=200$  nodes, $\beta=1.3$) and (bottom) the associated reduced~(condensed) lattice. The parameters $\eta,\zeta\geq 0$ specify the type of boundary conditions.
Top panel: Frequency dispersion bands and defect modes~(translation displacements). 
Bandgaps (light violet) and frequency range of negative mass (ochre) are also shown. Natural frequencies of the microstructure (red lines): $\omega_1^2=5.22,\omega_2^2=0.78$.}
\label{defectmode}
\end{figure}
%-----------------------------
%-----------------------------

%-----------------------------
%-----------------------------
\begin{figure}[t!]
\centering
\begin{subfigure}[b]{0.8\textwidth}
   \includegraphics[width=1\linewidth]{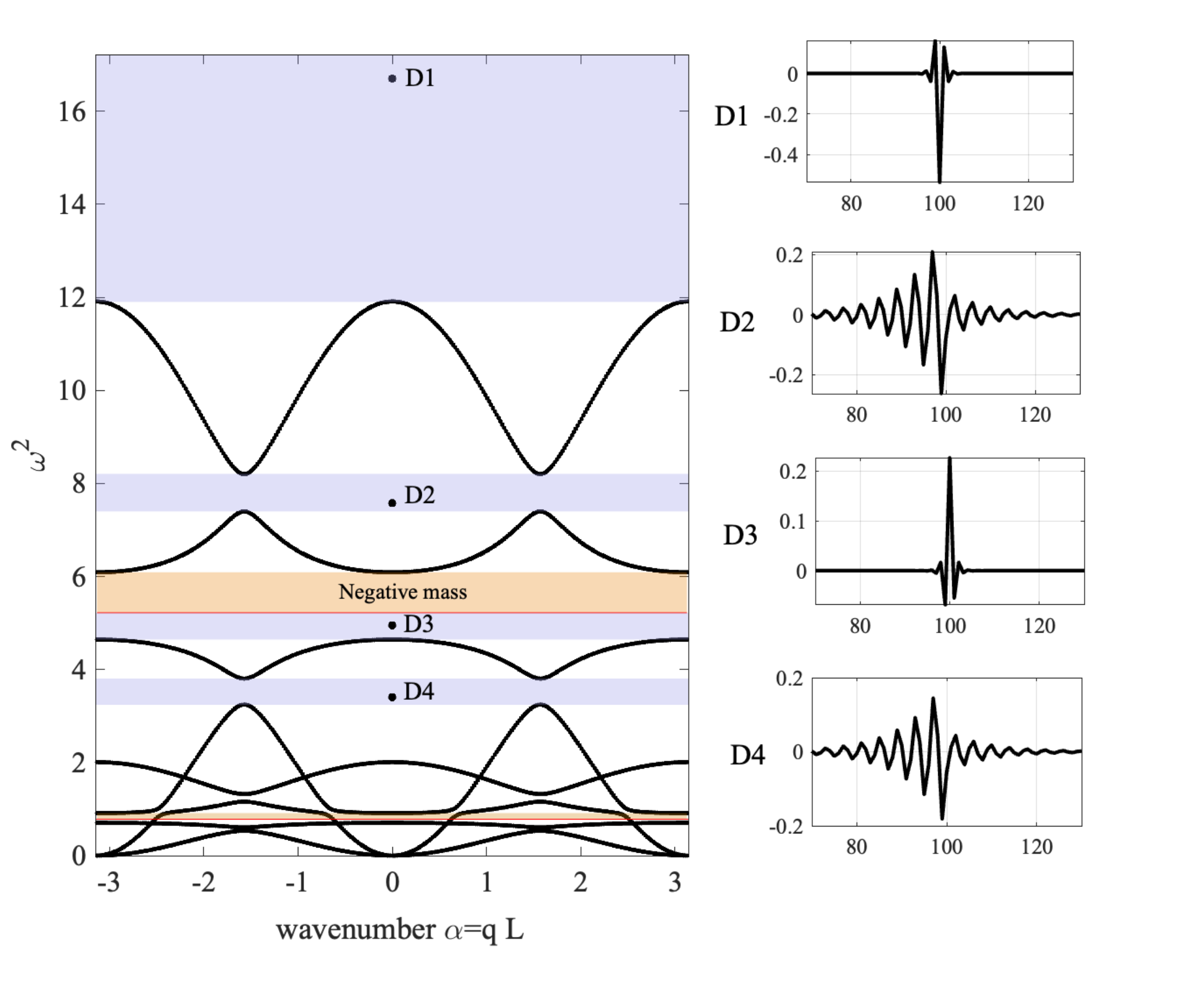}
\end{subfigure}
%-----------------------------
\vskip 0.3in
\begin{subfigure}[b]{0.9\textwidth}
   \includegraphics[width=1\linewidth]{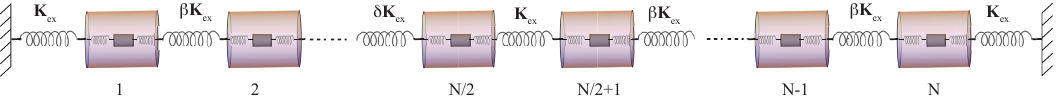}
\end{subfigure}
%-----------------------------
\vskip 0.2in
\begin{subfigure}[b]{0.9\textwidth}
   \includegraphics[width=1\linewidth]{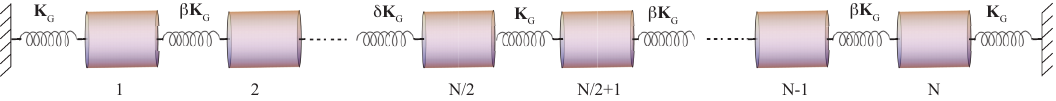}
\end{subfigure}
%-----------------------------
\caption{Bottom panel: (top) A $1$D defective finite composite lattice with alternating nearest-neighbor cell-cell stiffness matrices~($N=200$ nodes, $\beta=1.3$) and central defect of the outer, or macro stiffness~($\delta=4$); (bottom) the associated reduced~(condensed) lattice. 
Top panel: Frequency dispersion bands and defect modes~(translation displacements). 
Bandgaps (light violet) and frequency range of negative mass (ochre) are also shown. Natural frequencies of the microstructure (red lines): $\omega_1^2=5.22,\omega_2^2=0.78$.}
\label{defectmode2}
\end{figure}
%-----------------------------
%-----------------------------

%-----------------------------
%-----------------------------
\begin{figure}[t!]
\centering
\begin{subfigure}[b]{0.6\textwidth}
   \includegraphics[width=1\linewidth]{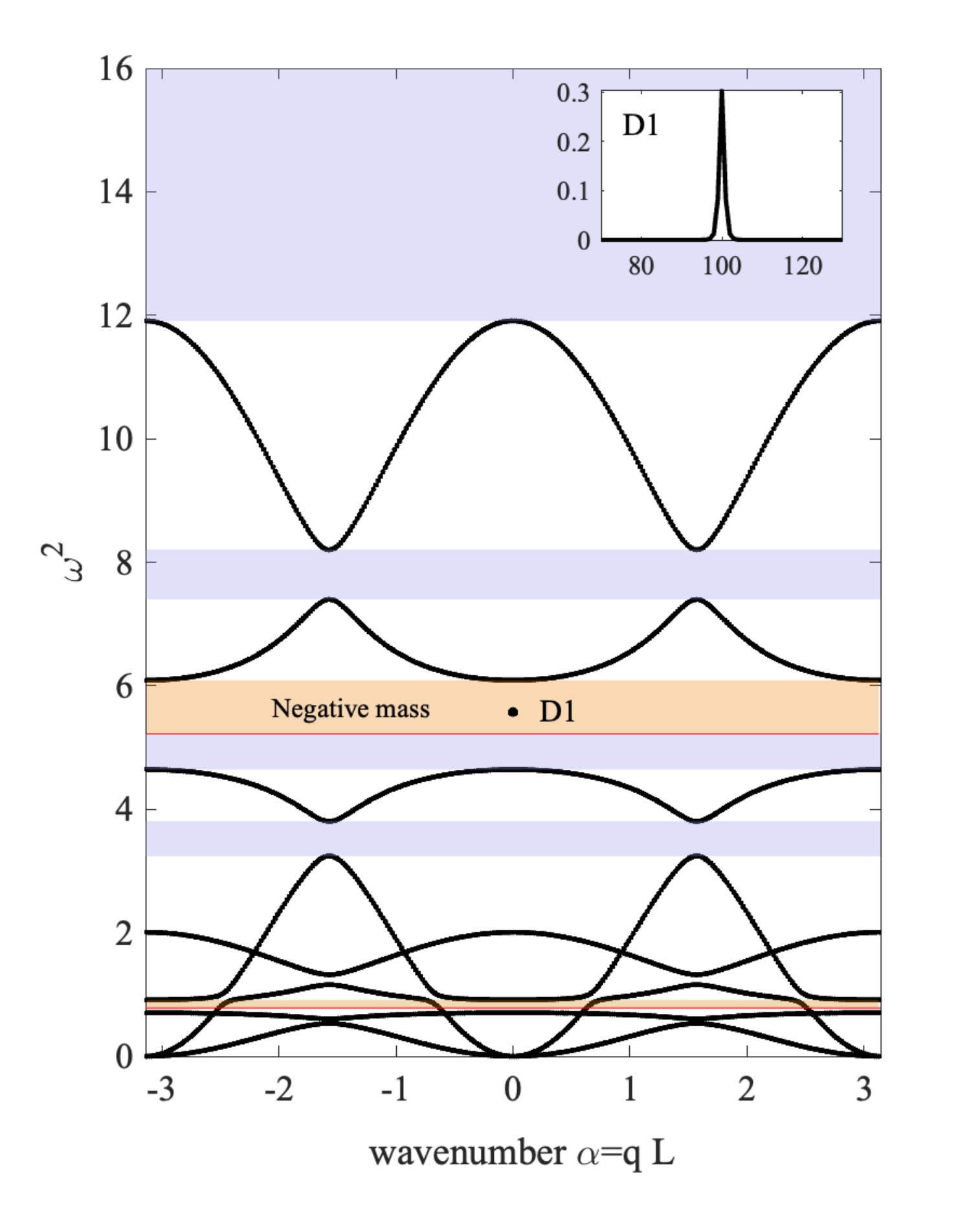}
\end{subfigure}
%-----------------------------
\vskip 0.3in
\begin{subfigure}[b]{0.9\textwidth}
   \includegraphics[width=1\linewidth]{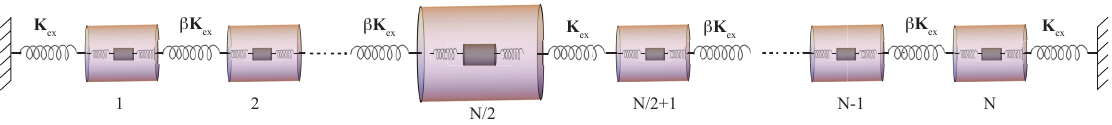}
\end{subfigure}
%-----------------------------
\vskip 0.2in
\begin{subfigure}[b]{0.9\textwidth}
   \includegraphics[width=1\linewidth]{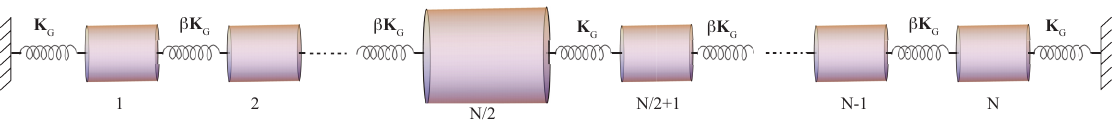}
\end{subfigure}
%-----------------------------
\caption{Bottom panel: (top) A $1$D defective finite composite lattice with alternating nearest-neighbor cell-cell stiffness matrices ($N=200$ nodes, $\beta=1.3$) and a microstructure defect at the central cell with micro mass $m$ and inertia $I_m$ $\gamma$-times larger than those of the other cells~($\gamma=1.75)$; (bottom) the associated reduced~(condensed) lattice. 
Top panel: Frequency dispersion bands 
and bandgaps (light violet) and frequency range of negative mass (ochre) are also shown. There is only defect mode~(translation displacements) of the defective lattice inside the negative-mass region of the bandgap.  Natural frequencies of the microstructure (red lines): $\omega_1^2=5.22,\omega_2^2=0.78$.}
\label{defectmode3}
\end{figure}
%-----------------------------
%-----------------------------

%-----------------------------
%-----------------------------
\subsection{Defects and localized modes}\label{Sec:defects}

Consider the defective $1$D composite lattice of $N$ cells indicated in the second panel from the bottom of Fig.~\ref{defectmode} with the local $4\times4$~stiffness matrix $\mathbf{K}_{\text{ex}}$ given in \eqref{KGs}$_1$, and the cell mass matrix $\mathbf{M}_{\text{cell}}$ given in \eqref{IOmatrixsys3}$_1$. The boundary springs $\eta \mathbf{K}_{\text{ex}}$ at $j=1$ and $\zeta \mathbf{K}_{\text{ex}}$ at $j=N$ (see bottom panel of Fig.~\ref{defectmode}) introduce defects as they break the translation symmetry of the lattice~(see, e.g., \cite{FedeleDefects2005}).

%------------------------------------------------------
%------------------------------------------------------
\subsubsection{The full lattice model}\label{Sec:fulllatticedefect}

The $4 N\times 4N$ global mass matrix is block-diagonal and is written as
%------------------------------------------------------
\begin{equation}\label{Globalmasss}
\mathbf{M}=\begin{bmatrix}
 \mathbf{M}_{\mathrm{cell}} &&&&\\
 &  \mathbf{M}_{\mathrm{cell}}&&&\\
 & & \ddots &&\\
 &  & & \mathbf{M}_{\mathrm{cell}} & \\
  &  & & &\mathbf{M}_{\mathrm{cell}}
\end{bmatrix}
\,.
\end{equation}
%------------------------------------------------------
The $4 N\times 4N$ global stiffness matrix is given by
%------------------------------------------------------
\begin{equation}\label{GlobalStiffsfinite}
\mathbf{K}=\begin{bmatrix}
\eta \mathbf{K}_{\text{ex}} + \beta\mathbf{K}_{\text{ex}}+\mathbf{K}_{\text{cell}} & -\beta \mathbf{K}_{\text{ex}} & \cdots & \cdots & \cdots\\
 -\beta \mathbf{K}_{\text{ex}} & (1+\beta)\mathbf{K}_{\text{ex}}+\mathbf{K}_{\text{cell}} & -\mathbf{K}_{\text{ex}} & \cdots & \cdots \\
 && \ddots &&\\
 && \ddots &&\\
 \cdots & \cdots & -\mathbf{K}_{\text{ex}} & (1+\beta)\mathbf{K}_{\text{ex}}+\mathbf{K}_{\text{cell}} & -\beta\mathbf{K}_{\text{ex}}\\
-\cdots &  \cdots & \cdots & -\beta\mathbf{K}_{\text{ex}} & \zeta \mathbf{K}_{\text{ex}}+\beta\mathbf{K}_{\text{ex}}+\mathbf{K}_{\text{cell}}
\end{bmatrix}\,.
\end{equation}
%------------------------------------------------------
The displacement vector has the form
$\hat{\mathbf{u}}=\left[\hat{\mathbf{u}}_{1}~\hat{\mathbf{u}}_{2}\cdots\hat{\mathbf{u}}_{N} \right]^{\mathsf{T}}$, where the vector $\hat{\mathbf{u}}_j\in \mathbb{C}^4$ lists the $2\times 1$ vectors of the Fourier amplitudes of the outer and inner generalized displacements $(\hat{\mathbf{X}}_{\text{O}},\hat{\mathbf{X}}_{\text{I}})$ of the unit cell $j$ at $X_j=j L$ given in \eqref{Fouriergendisp}.

%------------------------------------------------------
%------------------------------------------------------
\subsubsection{The reduced lattice model via the effective mass} 

We now consider the reduced lattice shown in the bottom panel of the same Fig.~\ref{defectmode}, where we lump the microstructure of each macro-mass to a single mass whose $2\times 2$~effective mass matrix is $\mathbf{M}_{\text{eff}}(\omega)$ given in~\eqref{1mass_effmass}. 
The $2\times2$~stiffness matrix of the elastic springs connecting the lumped masses is $\mathbf{K}_G$ given in \eqref{KGs}$_2$. 
The $2 N\times 2 N$ global effective mass matrix is block-diagonal and is written as
%------------------------------------------------------
\begin{equation}\label{Globalmasssdefect}
\mathbf{M}_{\text{e}}(\omega)=\begin{bmatrix}
 \mathbf{M}_{\mathrm{eff}}(\omega) &&&&\\
 &  \mathbf{M}_{\mathrm{eff}}(\omega)&&&\\
 & & \ddots &&\\
 &  & & \mathbf{M}_{\mathrm{eff}}(\omega) & \\
  &  & & &\mathbf{M}_{\mathrm{eff}}(\omega)
\end{bmatrix}
\,.
\end{equation}
%------------------------------------------------------
The $2 N\times 2 N$~global effective stiffness matrix is given by
%------------------------------------------------------
\begin{equation}\label{GlobalStiffsfinitedefect}
\mathbf{K}_{\text{e}}=\begin{bmatrix}
\eta \mathbf{K}_{\text{G}} + \beta\mathbf{K}_{\text{G}} & -\beta \mathbf{K}_{\text{G}} & \cdots & \cdots & \cdots\\
 -\beta \mathbf{K}_{\text{G}} & (1+\beta)\mathbf{K}_{\text{G}} & -\mathbf{K}_{\text{G}} & \cdots & \cdots \\
 && \ddots &&\\
 && \ddots &&\\
 \cdots & \cdots & -\mathbf{K}_{\text{G}} & (1+\beta)\mathbf{K}_{\text{G}} & -\beta\mathbf{K}_{\text{G}}\\
-\cdots &  \cdots & \cdots & -\beta\mathbf{K}_{\text{G}} & \zeta \mathbf{K}_{\text{G}}+\beta\mathbf{K}_{\text{G}}
\end{bmatrix}\,.
\end{equation}
%------------------------------------------------------
The displacement vector has the form
$\hat{\mathbf{v}}=\left[\hat{\mathbf{v}}_{1}~\hat{\mathbf{v}}_{2}\cdots\hat{\mathbf{v}}_{N} \right]^{\mathsf{T}}$, where $\hat{\mathbf{v}}_{j}$ is the $2\times 1$ vector of the Fourier amplitudes of the outer generalized displacements of the unit cell $j$ at $X_j=j L$, that is $\hat{\mathbf{v}}=\hat{\mathbf{X}}_{\text{O}}\in \mathbb{C}^2$  given in~\eqref{Fouriergendisp}$_1$. The matrix is Hermitian as expected.
Defect modes are identified by finding the isolated eigenvalues $\omega^2$ of the $2 N\times 2 N$~nonlinear eigenvalue problem $|\mathbf{K}_{\text{e}}-\omega^2\mathbf{M}_{\text{e}}(\omega)|=0$ by the bisection method~(see, e.g., \cite{FedeleDefects2005,FedeleDefects2005b} and references therein). The same modes can be obtained by solving the $4 N\times 4 N$~linear eigenvalue problem $|\mathbf{K}-\omega^2\mathbf{M}|=0$ of the full lattice, where the full mass and stiffness matrices are given in \S\ref{Sec:fulllatticedefect}. The defect eigenvalues reside lie the bandgaps of the periodic composite lattice.  

\paragraph{Examples}
The dispersion bands of the composite lattice~$(\beta=1.3,N=200)$ with an imperfection of the outer, or macro stiffness near node~$1$ are shown in Fig.~\ref{defectmode}. A defect mode is present in each bandgap of the associated periodic lattice. The defect modes are localized in space and tend to decay slower as they get closer to the band edges. Similar results hold for an imperfection of the outer, or macro stiffness at the central node of the lattice depicted in Fig.~\ref{defectmode2}. Note that in both cases the frequency of the defect mode $D_3$ is outside the bandgap region of negative mass. As a matter of fact, for the chosen parameters the microstructure has a positive-definite effective mass matrix. 
If one alters the microstructure of the central cell to have a negative effective mass, then the defective lattice has only one defect mode, which lies inside the negative-mass region of a bandgap as depicted in Fig.~\ref{defectmode3}.

%---------------------------
%------------------------------
\section{Concluding  Remarks} \label{Sec:Conclusions}

In this paper, we presented a general formalism for the effective mass of mechanical lattices with microstructure. Specifically, we first revisited a classical case of the microstructure being modeled as a spring-interconnected mass-in-mass system, showing that its frequency-dependent effective mass can be derived in three different ways, namely, momentum equivalence, action equivalence, and dynamic condensation of the momentum balance equations. Such an effective mass is the sum of a static mass and of an $\textit{added mass}$, which accounts for the effects of the microstructure on the macrostructure, in analogy to that of a swimmer in a fluid. This framework was generalized  to systems with arbitrary microstructure. 

As an application, we considered a $1$D composite lattice, whose microstructure is modeled by a chiral spring-interconnected mass-in-mass cell. A reduced (condensed) model of the full lattice is then obtained by lumping the microstructure into a single effective mass. We then studied the spectral properties of the composite lattice, in particular, the  frequency bands and localized modes due to defects. To do so, we performed a dynamic Bloch analysis using both the full and reduced lattice models, which provide identical spectral results. In particular, the frequency bands follow from the full lattice model by solving a linear eigenvalue problem, or from the reduced lattice model by solving a smaller nonlinear eigenvalue problem, for which we used the bisection method. We found that the range of frequencies of negative effective mass falls within the bandgaps of the lattice. In addition, localized modes due to defects of the macrostructure have frequencies within the bandgaps, but outside the negative-mass range. If the defect alters the microstructure of a lattice cell to have negative effective mass (micro-defect), then there is only one localized mode at the cell with defect, with frequency within the negative-mass range of the bandgap. Macro-defects of the outer lattice result in defect modes within each bandgap, but outside the negative-mass range, irrespective of the mass sign of the microstructure. 

The proposed formalism can be applied to reveal exotic properties of coupled micro-macro systems, such as active matter or metamaterials~\citep{Vitelli_topologicalmatter}, making it a worthy subject for future work. In particular, the unconventional properties of these  peculiar systems, such as negative mass or stiffness, and odd viscosity or elasticity~\citep{Vitelli_odd_elasticity}, are clearly defined for the reduced (or effective) lattice. 

%---------------------------
%------------------------------
\section*{Acknowledgement}

A.Y. was partially supported by NSF -- Grant No. CMMI 1939901, and  ARO Grant No. W911NF-18-1-0003. P.S. acknowledges support from the Clifford and William Greene, Jr. Professorship.

%--------------------------------------
%-----------------------------------------
%\bibliographystyle{plainnat}
\bibliography{ref,ref2,ref3,ref4}

\appendix

\end{document}